\documentclass[usenatbib]{mnras}
\usepackage{newtxtext,newtxmath}
\usepackage[T1]{fontenc}

\DeclareRobustCommand{\VAN}[3]{#2}
\let\VANthebibliography\thebibliography
\def\thebibliography{\DeclareRobustCommand{\VAN}[3]{##3}\VANthebibliography}
\usepackage{graphicx}	
\usepackage{amsmath}	
\usepackage{amssymb}	
\usepackage{multirow}
\usepackage{indentfirst}
\usepackage{stfloats}
\usepackage{url}

\newcommand{\be}{\begin{equation}}
\newcommand{\ee}{\end{equation}}
\newcommand{\ba}{\begin{eqnarray}}
\newcommand{\ea}{\end{eqnarray}}
\newcommand{\bi}{\begin{itemize}}
\newcommand{\ei}{\end{itemize}}

\title[Lensing magnification reconstruction]{A method of weak lensing reconstruction through cosmic magnification with multi-band photometry information}

\author[Ma et al.]{
Ruijie Ma,$^{1,2}$ Pengjie Zhang,$^{1,3,2}$\thanks{E-mail:zhangpj@sjtu.edu.cn}
Yu Yu,$^{1,2}$\thanks{E-mail:yuyu22@sjtu.edu.cn} Jian Qin,$^{1,2}$
\\
$^1$Department of Astronomy, School of Physics and Astronomy, Shanghai
Jiao Tong University, Shanghai, 200240, China\\
$^2$Key Laboratory for Particle Astrophysics and Cosmology
(MOE)/Shanghai Key Laboratory for Particle Physics and Cosmology,
China\\
$^3$Tsung-Dao Lee Institute, Shanghai Jiao Tong University , Shanghai
200240, China\\
}

\date{Accepted XXX. Received YYY; in original form ZZZ}

\pubyear{2021}

\begin{document}
\label{firstpage}
\pagerange{\pageref{firstpage}--\pageref{lastpage}}
\maketitle

\begin{abstract}
 Weak gravitational lensing induces flux dependent fluctuations in the observed galaxy number density distribution. This cosmic magnification (magnification bias) effect in principle enables lensing reconstruction alternative to cosmic shear and CMB lensing. However, the intrinsic galaxy clustering, which otherwise overwhelms the signal, has hindered its application. Through a scaling relation found by principal component analysis of the galaxy clustering in multi-band photometry space, we design a minimum variance linear estimator to  suppress the intrinsic galaxy clustering and to reconstruct the lensing convergence map. In combination of the CosmoDC2 galaxy mock and the CosmicGrowth simulation, we test this proposal for a LSST-like galaxy survey with $ugrizY$ photometry bands. (1) The scaling relation holds excellently at multipole $\ell<10^3$, and remains reasonably well to $\ell\sim 3000$. (2) The linear estimator efficiently suppresses the galaxy intrinsic clustering, by a factor of $\sim 10^2$. (3) For galaxies in the photo-z range $0.8<z_\kappa<1.2$, the reconstructed convergence map is cosmic variance limited per $\ell$ mode at $\ell<10^2$, and shot noise limited at $\ell\ga 200$.  (4) Its cross-correlation with cosmic shear of galaxies can achieve $S/N\ga 200$. When the source redshift of cosmic shear galaxies $z_\gamma<z_\kappa$, the systematic error is negligible at all investigated scales ($\ell<3000$). When $z_\gamma\geq z_\kappa$, the systematic error caused by the residual intrinsic galaxy clustering becomes non-negligible. We discuss possible mitigation of the residual intrinsic galaxy clustering  required for accurate measurement at $\ell>10^3$. This work further demonstrates the potential of lensing measurement through cosmic magnification to enhance the weak lensing cosmology.

\end{abstract}



\section{Introduction}
\label{sec:introduction}
Weak gravitational lensing is a major probe of dark matter, dark energy, gravity, cosmic neutrinos and the origin mechanism of the Universe \citep{2003ARA&A..41..645R,DETF,Hoekstra08,LSSTDE12,Weinberg13}. After decades of efforts, weak lensing measurements through the induced distortion in the cosmic microwave background  temperature and polarization maps (CMB lensing, \citet{2020A&A...641A...8P,2023arXiv230405202Q}) and galaxy shapes (cosmic shear, \citet{2022PhRvD.105b3514A,2023arXiv230400702L}) have both achieved $\mathrm{S/N}\gtrsim 40$. Upcoming stage IV dark energy surveys (e.g. CSST \citep{2019ApJ...883..203G,2023arXiv230404489Y}, Euclid \citep{Euclid16}, and LSST \citep{LSST,2018RPPh...81f6901Z}) and  future CMB experiments such as CMB-S4 \citep{CMBS4} will improve the S/N to a level of $\gg 10^2$ .

Weak lensing also magnifies/amplifies galaxy flux/size (e.g. \citet{2011PhRvD..84j3004V,2012ApJ...744L..22S}) and induces coherent distortion in the galaxy number density distribution (magnification bias, e.g. \citet{2002ApJ...580L...3J}). For the magnification bias, the lensing signal is overwhelmed by the intrinsic galaxy clustering. In contrast to the largely uncorrelated galaxy shape noise in cosmic shear measurement, the intrinsic galaxy clustering has strong spatial correlation and can not be simply averaged out. There are two existing approaches to deal with this issue. The conventional method   cross-correlates the number density distribution of two galaxy/quasar samples with non-overlapping redshift distribution. Since the intrinsic galaxy clustering vanishes for such large spatial separation, the measured cross-correlation is dominated by the foreground galaxy density-background galaxy magnification bias cross-correlation. The first detection using galaxies as lenses was achieved through cross-correlating high redshift SDSS quasars with low redshift SDSS luminous red galaxies (LRGs), so was the characteristic dependence on the galaxy flux \citep{2005ApJ...633..589S}. A joint measurement and analysis together with dust extinction were performed by \citet{2010MNRAS.405.1025M}. Later detections have been extended to  different redshift ranges and/or different high redshift tracers, such as Lyman-break galaxies at $2.5<z<5$ \citep{2009A&A...507..683H,2013MNRAS.429.3230H}, submillimetre galaxies at $z\sim 2$ \citep{2011MNRAS.414..596W,2014MNRAS.442.2680G,2021A&A...656A..99B,2021A&A...646A.152G} and normal galaxies in the DES survey \citep{2018MNRAS.476.1071G}.  Such method is successful and is the standard approach to measure the magnification bias. Refer to \cite{2020PhRvL.124c1101J,2021MNRAS.504.1452V,2022A&A...662A..93E,2022arXiv220909782E} for recent investigations on cosmic magnification and its cosmological impacts.

However the measured cross-correlation depends on the (foreground/lens) galaxy bias, which is hard to predict from the first principle and therefore reduces its cosmological constraining power. There are possibilities to circumvent the problem of galaxy bias. One possibility is to use the galaxy auto-correlation measurement for constraining the galaxy bias. This is accurate at sufficiently large scale. But at small scale a further issue to take care of is the galaxy-matter stochasticity (galaxy-matter cross-correlation coefficient $r\neq 1$). Another possibility recently proposed by  \citet{2021PhRvD.103l3504L} and implemented in the HSC data is to measure the cosmic magnification-cosmic shear cross-correlation. Being immune to the galaxy bias, the above cross-correlation measurement is attractive for the purpose of cosmological constraints.

 An alternative approach of cosmic magnification measurement is to eliminate the intrinsic galaxy clustering through the characteristic dependence of cosmic magnification (magnification bias) on galaxy flux \citep{Zhang05}. This idea was implemented in \citet{YangXJ11,YangXJ15}, under the limit of deterministic galaxy bias. However, galaxy bias is known to have stochastic components (e.g. \citet{Bonoli09,Hamaus10,Baldauf10}). To deal with such stochasticity, an analytical method of blind separation (ABS) was proposed in the context of CMB foreground removal \citep{ABS}, and applied to cosmic magnification \citep{YangXJ17,Zhang18}. This method is mathematically rigorous, as long as the number of eigenmodes of intrinsic galaxy clustering is smaller than the number of flux bins. However, the price to pay for the blind separation is a significantly weaker S/N than cosmic shear \citep{Zhang18}. To improve the S/N,  \citet{2021RAA....21..247H} proposed a constrained internal linear combination (cILC) method. This method  eliminates the mean galaxy bias, but not its variation across galaxy sub-samples.

Here we design a new method of weak lensing reconstruction through cosmic magnification, with two improvements over our previous method of \citet{2021RAA....21..247H}. (1) It utilizes a scaling relation in galaxy clustering to suppress the intrinsic galaxy clustering. This scaling relation has been verified both in this paper for the case of 2D angular clustering, and in \citet{2023arXiv230411540Z} for the case of 3D clustering. This enables a more efficient suppression of the deterministic galaxy bias. (2) It also utilizes multiple photometry band information available in many imaging surveys of galaxies such as CSST and LSST. Twofold benefits can be achieved. One is the improved accuracy of the above scaling relation and the resulting improvement in suppressing the deterministic bias. The other is the suppression of shot noise, since the relative ranking of galaxy brightness varies across photometry bands.

The paper is organized as follows. \S \ref{sec:method} introduces the method of lensing reconstruction through cosmic magnification. \S \ref{sec:performance} shows the performance test with mocks. Details of the mocks are described in \S \ref{sec:mocks}. Further details on the performance test results are shown in \S \ref{sec:errors}. \S \ref{sec:conclusion} summarizes this work and discuss remaining issues.

\begin{table}	
\centering
\begin{tabular}{c|cccccccccc}
\hline
	\hline
	Photometry band  & $u$ & $g$ & $r$ & $i$ & $z$ & $Y$ \\
    Magnitude cut & 26.0 & 27.3 & 27.4 & 26.6 & 26.1 & 24.9\\
    $\bar{n}$ of each band(${\rm arcmin}^{-2}$) & 1.27 & 9.14 & 13.97 & 9.84 & 7.28 & 2.04 \\
    Number of flux bins & 2 & 5 & 7 & 6 & 4 & 2 \\

	\hline
	\end{tabular}	
	\caption{The galaxy mocks to test our method. They are generated using the $1200$ Mpc$/h$ box size and $3072^3$ simulation particle CosmicGrowth simulation, based on the galaxy-halo relation of the  CosmoDC2 mock galaxies (details in the appendix \ref{sec:mocks}). The $ugrizY$ magnitude cut corresponds to a LSST/Rubin observatory-like survey.  Unless otherwise specified, we only consider galaxies in the photo-z range  $0.8<z_\kappa<1.2$. To avoid confusion, throughout the paper we use $z_\kappa$ to denote the (photometric) redshift of galaxies used for constructing the lensing convergence $\kappa$ through cosmic magnification, and use $z_\gamma$ for redshift of galaxies for cosmic shear measurements. $\bar{n}$ is the mean galaxy number density at $0.8<z_\kappa<1.2$. For each photometry band, we split galaxies into flux bins with equal number of galaxies. We use  26 galaxy sub-samples in total.  \label{tab:band}}	
\end{table}
\section{The map reconstruction method}
\label{sec:method}
Gravitational lensing changes the galaxy number overdensity to (e.g. \citet{2002ApJ...580L...3J})
\ba
\label{eqn:deltaL}
\delta_g^L=\delta_g+q\kappa\ .
\ea
Here $\delta_g$ is the intrinsic galaxy number overdensity. The above formula applies both to 3D space and 2D space. Hereafter we focus on the 2D case, in which $\delta_g$ denotes the surface overdensity. $\kappa$ is the lensing convergence. The coefficient $q$ is the response of $\delta_g^L$ to weak lensing. Given the galaxy selection criteria and observational conditions, $q$ can be robustly inferred (e.g. recent works by  \citet{2021MNRAS.504.1452V,2022arXiv220909782E}). For ideal case of flux limited sample, $q=2(\alpha-1)$ and $\alpha$ is the logarithmic slope of galaxy luminosity function. For brevity, our analysis adopts this simplification. $\delta_g$ can be decomposed as \citep{YangXJ17}
\ba
\label{eqn:deltag}
\delta_g=\delta_g^D+\delta_g^S\ \ .
\ea
Here $\delta_g^S$ is the stochastic component in the sense that its two-point correlation with the matter overdensity $\delta_m$ vanishes ($\langle \delta_g^S\delta_m\rangle=0$). $\delta_g^D$ is the deterministic component and $b_D$ is the corresponding deterministic galaxy bias,
\begin{equation}
\label{eqn:deltagD}
\delta_g^D\equiv b_D \delta_m\ ,\   b_D(\ell)=\frac{C_{gm}(\ell)}{C_{mm}(\ell)}\ .
\end{equation}
$C_{mm}$ and $C_{gm}$ are the matter auto power spectrum and the galaxy-matter cross power spectrum, respectively. Notice that $b_D(\ell)$ is in general scale dependent. When $\delta_g^S\neq 0$, $b_D$ differs from the other definitio of galaxy bias $\sqrt{C_{gg}/C_{mm}}$.  $\delta_g^S$ mainly contaminates the lensing auto-correlation, but $\delta_g^D$ contaminates both the lensing auto-correlation and cross-correlations with other large scale structure (LSS) fields.

The primary goal of our method is to eliminate $\delta_g^D$ in the reconstructed map, such that its cross-correlations with other LSS fields (e.g. cosmic shear \citep{2021PhRvD.103l3504L}) are unbiased. For a photo-z bin, we split galaxies into sub-samples according to their flux in a photometry band (e.g. $r$ band). We then repeat for other photometry bands. Therefore galaxies in different sub-samples may overlap.  We denote $\bar{n}_\alpha$ as the average number density of galaxies in the $\alpha$-th sub-sample, and $\bar{n}_{\alpha\beta}$ as the average number density of galaxies in both the $\alpha$-th and $\beta$-th sub-samples.  We denote $\delta_{g,\alpha}(\vec{\theta})$ (and $\delta_{g,\alpha}(\vec{\ell})$ in Fourier space) as the overdensity of the $\alpha$-th sub-sample. Here $\alpha=1,\cdots, N$ and $N$ is the total number of sub-samples.

Without loss of generality, hereafter we will adopt the flat sky approximation and  work in the Fourier space. We adopt a linear estimator to estimate $\kappa(\vec{\ell})$,
\begin{equation}
\label{eqn:kappahat0}
    \hat{\kappa}(\vec{\ell})=\sum_\alpha W_\alpha(\ell) \delta_{g,\alpha}^L(\vec{\ell})\ .
\end{equation}
Here $\vec{\ell}$ is the 2D wavevector and $\ell$ is its amplitude.
$W_\alpha$, depending on both $\alpha$ and $\ell$, is a weight to be determined. Ideally it should satisfy $\sum_\alpha W_\alpha b_{D,\alpha}=0$. But the problem is that $b_{D,\alpha}$ is not a direct observable. Recently we confirmed a proportionality relation between $b_{D,\alpha}$ and a direct observable \citep{2023arXiv230411540Z},
\begin{equation}
\label{eqn:be}
    b_{D,\alpha}\propto E_\alpha^{(1)}\ .
\end{equation}
This relation was originally postulated by \cite{1999ApJ...518L..69T}. \citet{2023arXiv230411540Z} verified it using the IllustrisTNG simulations.
From the $N$ galaxy sub-samples, we can measure their cross power spectra $C_{\alpha\beta}(\ell)$).
$E_\alpha^{(1)}$ is the eigenvector of the largest eigenmode of this $N\times N$ matrix.
This relation is nearly exact in 3D space as we have verified using TNG simulated galaxies, with the  cross-correlation coefficient $r\simeq 1$ ($1-r^2\la 10^{-4}$).
\begin{figure}
\centering
  \includegraphics[width=0.45\textwidth]{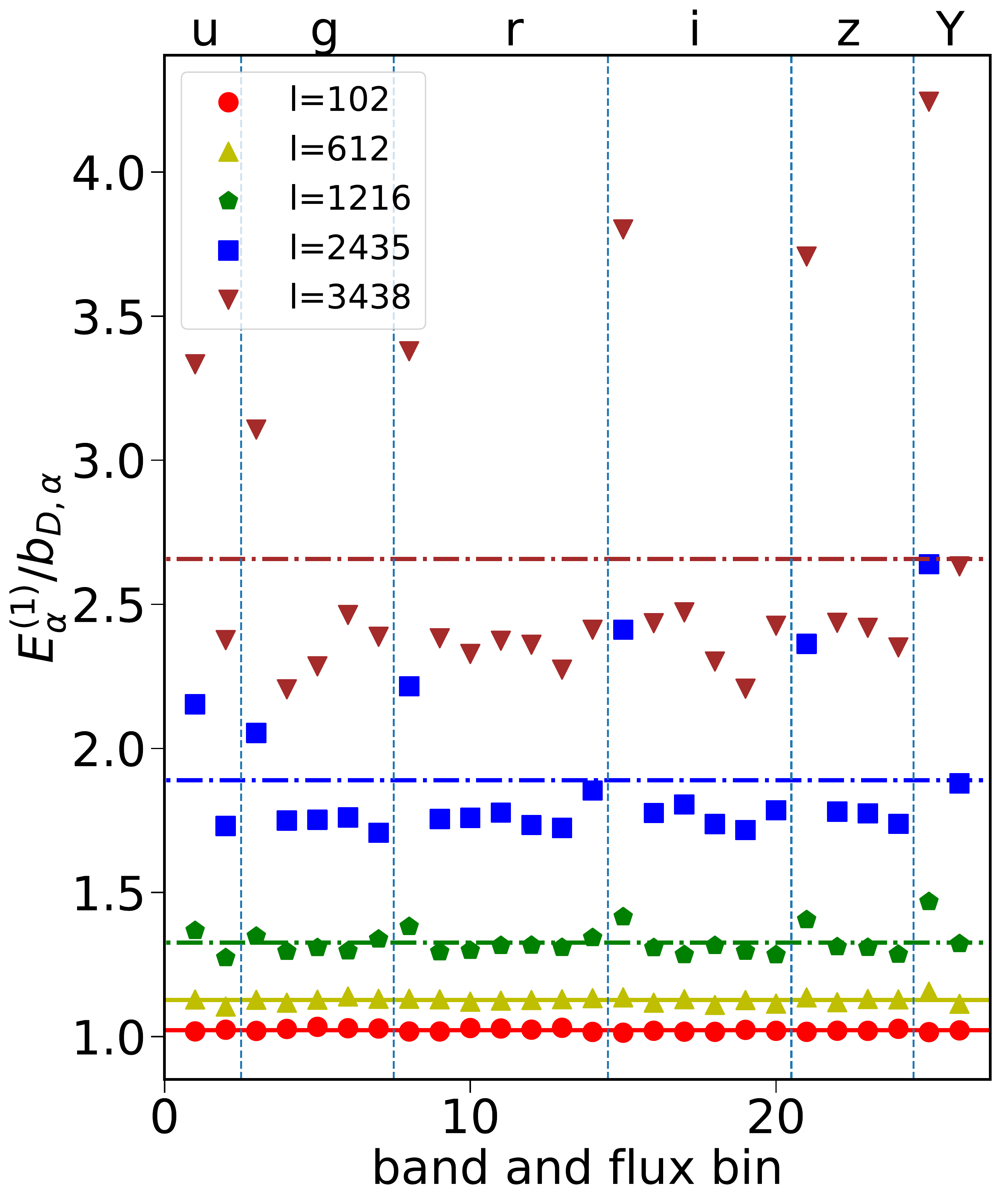}
	\caption{The validity of the scaling relation Eq. \ref{eqn:be}. We plot the ratio $E_\alpha^{(1)}/b_{D,\alpha}$ at various $\ell$. $\alpha=1,\cdots,26$ denotes galaxy sub-samples defined in Table \ref{tab:band}. The accuracy of the  proportionality relation between $E^{(1)}_\alpha$ and $b_{D,\alpha}$ is excellent at $\ell<10^3$, and is reasonable at $\ell<3000$.
	\label{fig:eme}}
\end{figure}

The relation is less accurate  in photometric redshift space. Due to significant photo-z uncertainties, the true redshift distribution of photo-z galaxy samples is inevitably wide. This mixes different $k$ scale and galaxy intrinsic properties.  We test the validity of Eq. \ref{eqn:be} in the 2D angular clustering against CosmoDC2. CosmoDC2 is the galaxy mock generated for LSST \citep{CosmoDC2}. It provides a lightcone of $440\ \mathrm{deg}^2$, with $\sim 2.26$ billion galaxies to $z=3$. The magnitudes in $ugrizY$ photometry bands  are also provided. Such  multi-band photometry information is crucial for our method. Since CosmoDC2 does not provide the underlying dark matter density light cones, we map CosmoDC2 galaxies to a CosmicGrowth simulation \citep{Jing19}, based on halo abundance matching. We then randomly add photo-z errors of $\sigma_z=0.05(1+z)$ to mock galaxies, and generate 20 lightcones, each of size $8.2^{\circ}\times 8.2^{\circ}$. Details of the simulations and the mocks are described in \S \ref{sec:mocks}.

Table \ref{tab:band} shows the definition of galaxy sub-samples in $ugrizY$ photometry bands and $26$ flux bins in total ($2,5,7,6,4,2$ flux bins in  $ugrizY$ bands respectively).  Unless otherwise specified, these will be the default galaxy samples for tests in this paper. We have investigated multipole photometric redshift bins for lensing reconstruction. Since the method is found to be  applicable in all cases, we only show the default photometric redshift range $0.8<z_\kappa<1.2$, for brevity. The number density of galaxies, ranging from $1.3$ arcmin$^{-2}$ to $14$ arcmin$^{-2}$, varies significantly with the photometry band (and the associated magnitude cut).

Fig. \ref{fig:eme} shows $E_\alpha^{(1)}/b_{D,\alpha}$ for the above galaxy sub-samples. The proportionality relation (Eq. \ref{eqn:be}) is nearly exact at $\ell\la 1000$ and remains reasonably valid  until $\ell\sim 3000$ The accuracy of the proportionality relation also depends on the flux bins.  The bins that deviates the most from the proportionality relation are the brightest galaxies in the given photometry bands (e.g. the uppermost 6 data points of Fig. 1). Therefore the behavior may be related to the nonlinear bias of these brightest galaxies. One issue is whether or not to use these bins in the cosmic magnification reconstruction. Disregarding them indeed reduces the systematic bias due to better proportionality relation. But on the other hand,  it enhances the statistical errors. The reason is that the magnification coefficient $q$ in these bins are significantly different to other bins. Without these bins, the weighting scheme designed later is less efficient and the reconstruction uncertainty is enhanced. The combined consequence is that disregarding them does not improve the overall reconstruction performance.  Therefore for the rest of the paper, we will include these bins in the analysis. Nevertheless, this issue deserves further investigation, in mock data and in real data analysis.

We can then determine $W_\alpha$ by the following three conditions,
\begin{equation}
\label{eqn:wq}
    {\rm Vanishing\  multiplicative\  error:} \sum_\alpha W_\alpha q_\alpha=1\ ,
\end{equation}
\begin{equation}
\label{eqn:we}
{\rm Eliminating}\ b_D:    \sum_\alpha W_\alpha E_\alpha^{(1)}=0\ ,
\end{equation}
\begin{equation}
\label{eqn:ws}
    {\rm minimizing\ shot\ noise:}\  N_{\rm shot}=\sum_{\alpha\beta} W_\alpha W_\beta \frac{\bar{n}_{\alpha\beta}}{\bar{n}_\alpha \bar{n}_\beta}\ .
\end{equation}
 The first condition (Eq. \ref{eqn:wq}) is to make the estimation $\hat{\kappa}$ free of multiplicative bias, given the magnification response parameter $q_\alpha$. The second (Eq. \ref{eqn:we}) is to eliminate the deterministic galaxy bias. The third is to minimize the shot noise.  The three conditions fix
 \begin{equation}
 \label{eqn:W}
     {\bf W}=\frac{C {\bf s}^{-1} {\bf q}-B{\bf s}^{-1} {\bf E}}{AC-B^2}\ .
 \end{equation}
 Here
 $A\equiv {\bf q}^T{\bf s}^{-1}{\bf q}$, $B\equiv {\bf q}^T{\bf s}^{-1}{\bf E}$ and  $C\equiv {\bf E}^T{\bf s}^{-1}{\bf E}$. $s_{\alpha\beta}\equiv \bar{n}_{\alpha\beta}/\bar{n}_\alpha\bar{n}_\beta$ is the shot noise matrix.  The method is identical to that in  \citet{2021RAA....21..247H} proposed by some of the authors, but with two important improvements by replacing the approximate condition $\sum_\alpha W_\alpha=0$ \citep{2021RAA....21..247H} with Eq. \ref{eqn:we}, and by including multiple photometry band information.

\begin{figure}
    \includegraphics[width=0.24\textwidth]{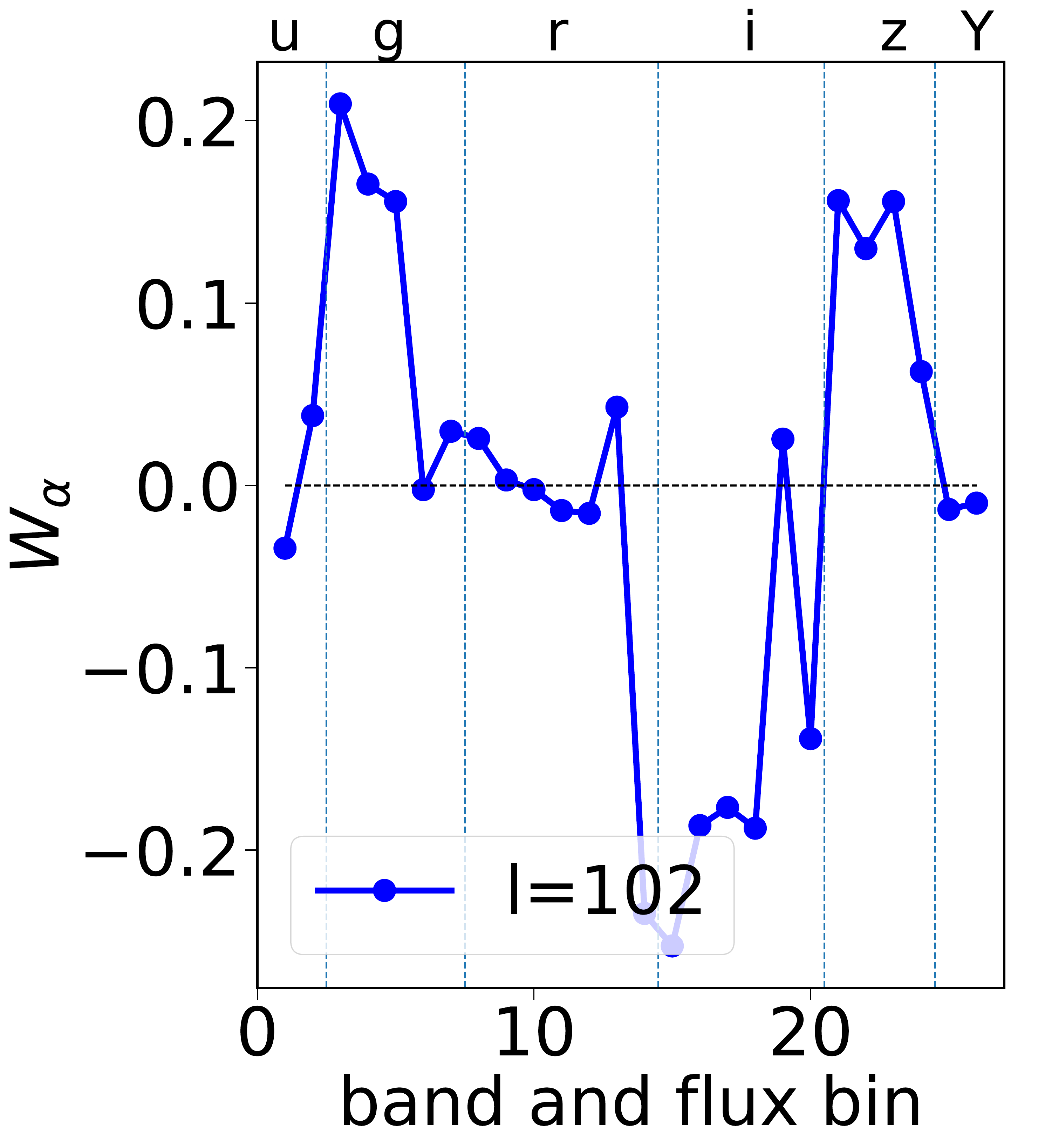}
    \includegraphics[width=0.24\textwidth]{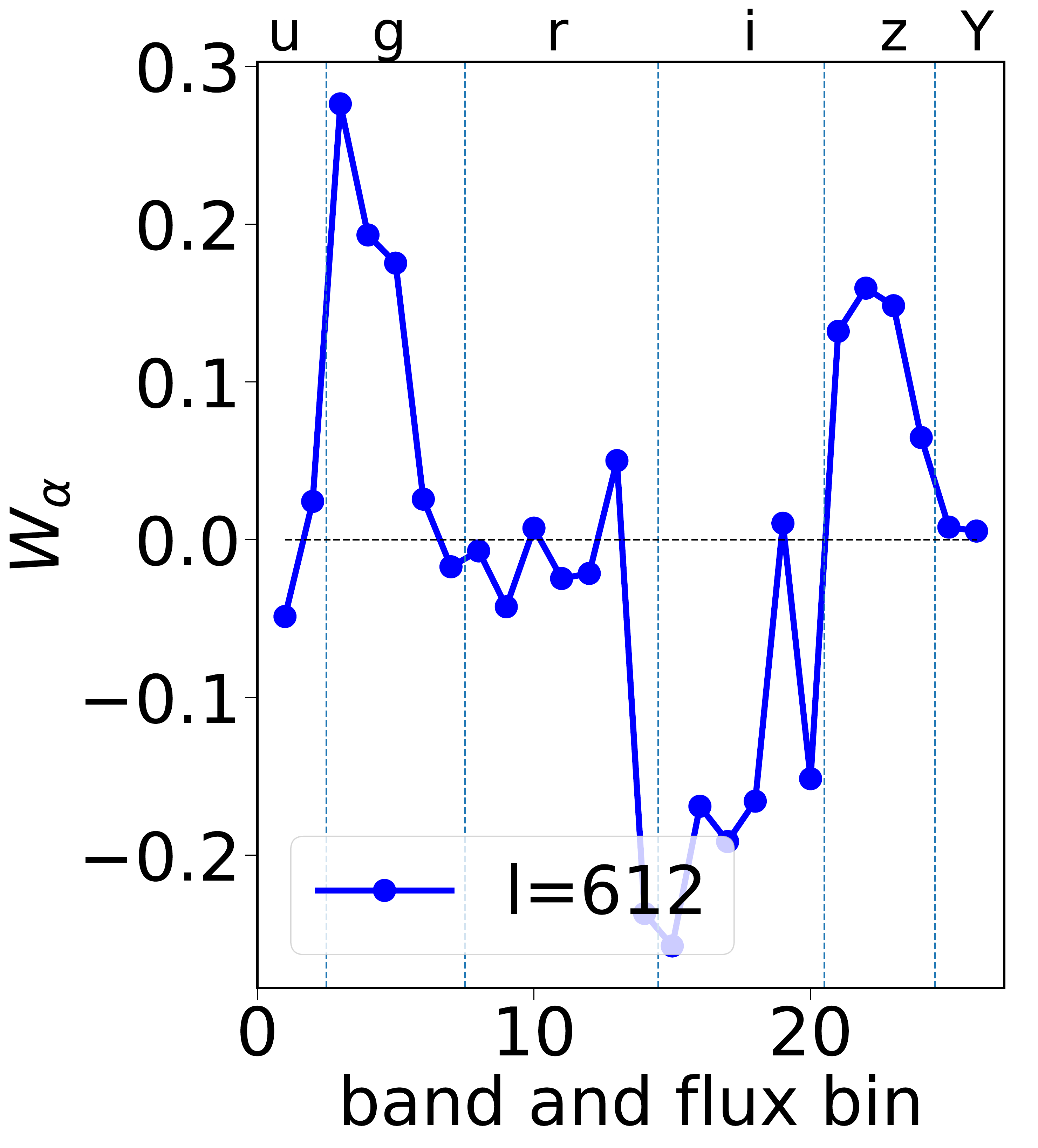}
    \includegraphics[width=0.24\textwidth]{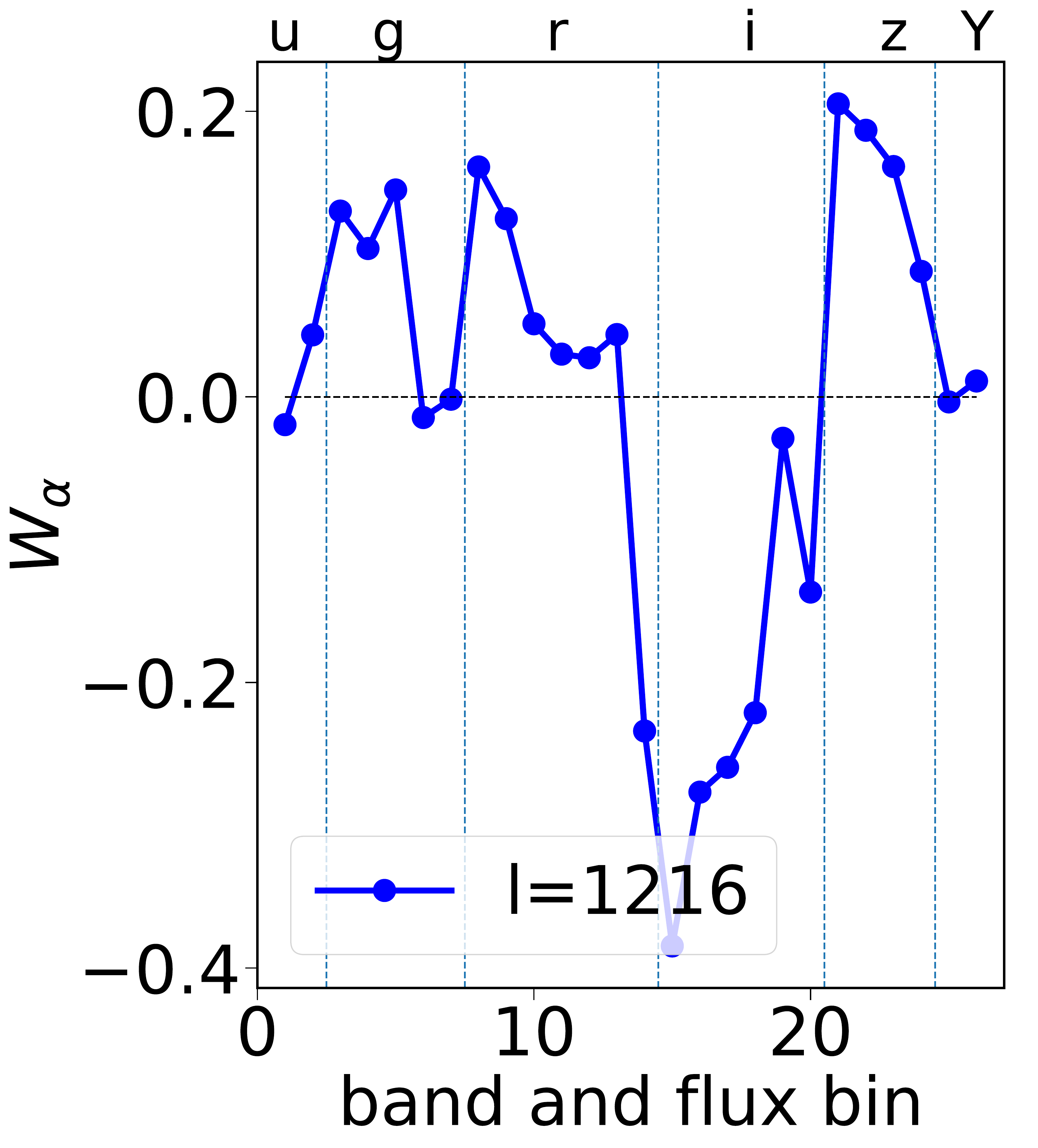}
    \includegraphics[width=0.24\textwidth]{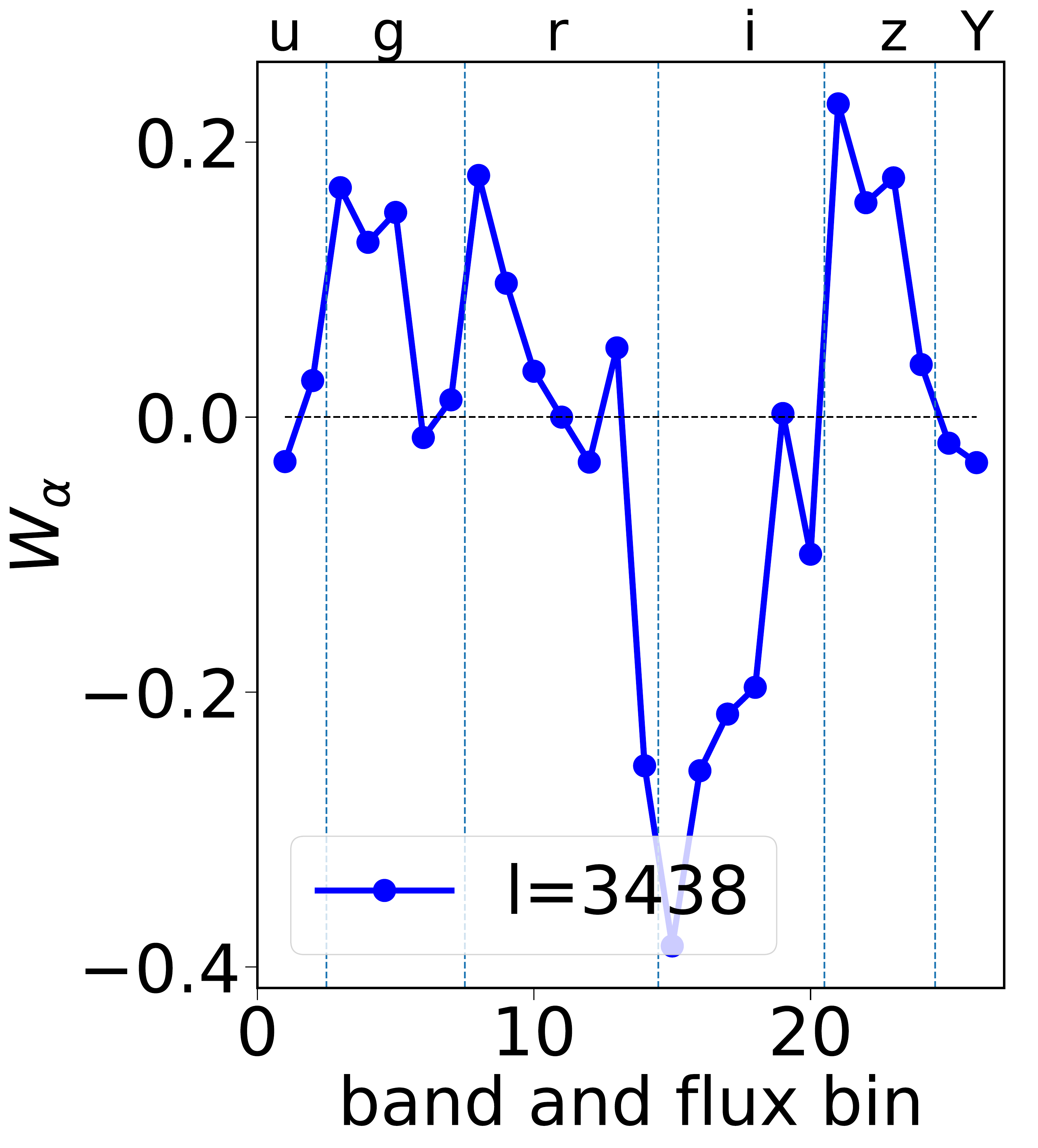}
	\caption{The weight $W_\alpha$ of the 8th map, which is chosen randomly among the 20 maps. We plot the weight at 4 angular scales ($\ell=102,612,1216,3438$). $\alpha=1,\cdots 26$ denotes the $\alpha$-th galaxy sub-sample defined by the corresponding photometry band and flux bin (Table \ref{tab:band}).  \label{fig:weight}}
\end{figure}

Fig. \ref{fig:weight} shows the weight $W_\alpha$ at various multipole $\ell$. The dependence on $\ell$ is visible. This is induced by the condition Eq. \ref{eqn:we} through $E_\alpha^{(1)}(\ell)$. $W_\alpha$ also changes sign. This is also the consequence of the condition Eq. \ref{eqn:we}. Interestingly, the mean $W$ is a factor of $10$ than the typical value of $|W_\alpha|\sim 0.1$ (Fig. \ref{fig:sumwi}). So the approximation $\sum_\alpha W_\alpha=0$ adopted by \citet{2021RAA....21..247H} is roughly satisfied. When we are not able to infer $E^{(1)}_\alpha$ robustly from observations (e.g. with the presence of uncorrected selection effects affecting the clustering measurement), the method of  \citet{2021RAA....21..247H} will be a viable alternative.

\begin{figure}
    \centering
   \includegraphics[width=0.45\textwidth]{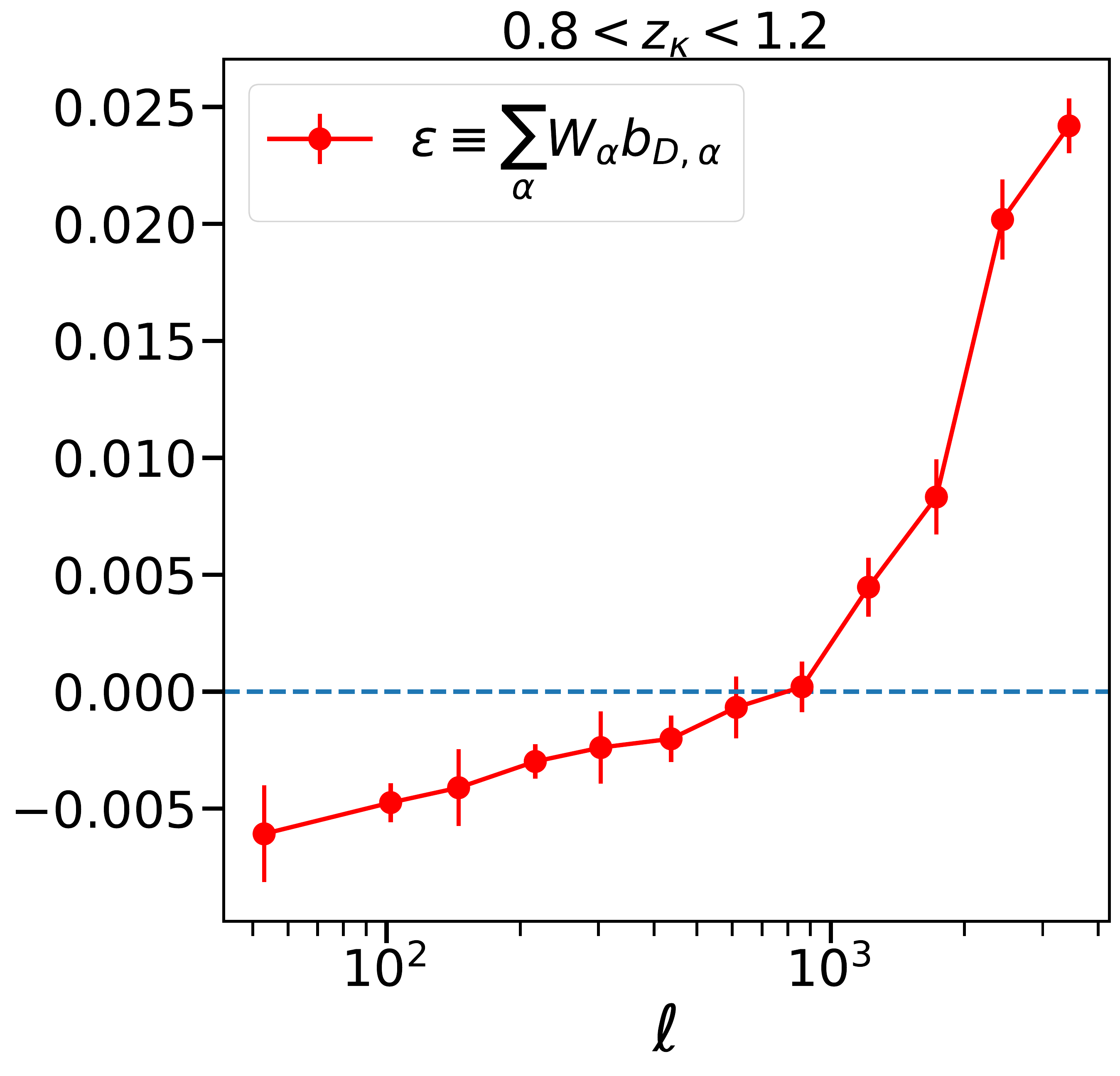}
    \caption{Multiplicative error in the reconstruction, proportional to $\epsilon\equiv \sum_\alpha W_\alpha b_{D,\alpha}$. The weighting scheme efficiently suppresses the galaxy intrinsic clustering by a factor of $\sim 10^2$. But since the proportionality relation (Eq. \ref{eqn:be}) is not exact, $\epsilon\neq 0$ and there will be residual multiplicative error at nonlinear scales ($\ell\ga 2000$). On the other hand, $\epsilon<0$ at $\ell\la 10^2$ is likely caused by the matrix operation, which renders statistical errors in $C_{\alpha\beta}$ into systematic errors in $E^{(1)}_\alpha$. }
    \label{fig:epsilon}
\end{figure}

\section{The performance}
\label{sec:performance}
One of the reconstructed maps is shown in Fig. \ref{fig:kappafield}.  The maps are noisy due to overwhelming shot noise. Therefore we have to quantify the reconstruction performance statistically.  Other than shot noise fluctuation, the reconstructed  map is
\begin{equation}
\label{eqn:kappahat}
    \hat{\kappa}=\kappa+\epsilon \delta_m+\sum_\alpha W_\alpha \delta_{g,\alpha}^S\ .
\end{equation}
The above relation comes from Eq. \ref{eqn:kappahat0}, together with Eq. \ref{eqn:deltaL}, \ref{eqn:deltag} \& \ref{eqn:deltagD}.
The error term $\sum W_\alpha \delta_{g,\alpha}^S$  is caused by stochastic galaxy bias. It can be down-weighted, but not eliminated by our method. This term is uncorrelated with the matter overdensity $\delta_m$ in the two-point correlation. So it does not induce systematic errors in cross-correlation with external weak lensing data sets, such as cosmic shear and CMB lensing. It contaminates the auto-correlation measurement. But it is sub-dominant comparing to the shot noise fluctuation from discrete galaxy distribution.

\subsection{Residual deterministic bias}
The $\epsilon \delta_m$ term is analogous to the galaxy intrinsic alignment in the cosmic shear measurement. It contaminates both the auto and cross power spectrum measurements.

Since the proportionality relation (Eq. \ref{eqn:be}) is not completely exact and since in reality $C_{\alpha\beta}$  suffers from shot noise, cosmic variance, magnification bias and other observational effects, the condition Eq. \ref{eqn:we} does not guarantee $100\%$ removal of deterministic galaxy clustering. The residual error in the $\kappa$ reconstruction is $\epsilon\delta_m$. Here
\begin{equation}
\epsilon\equiv \sum W_\alpha b_{D,\alpha}\neq 0\ .
\end{equation}
Fig. \ref{fig:epsilon} shows that $|\epsilon|\la 5\times 10^{-3}$ at $\ell<1000$, but increases to $0.01$-$0.03$ at $\ell\in (1000,4000)$. The negative $\epsilon$ at $\ell\la 100$ is surprising, since naively we may expect that at such large scale the galaxy stochasticity vanishes and $\epsilon=0$. This may be caused by statistical fluctuations in $C_{\alpha\beta}$ due to limited number of multipole $\ell$ modes. These statistical fluctuations in $C_{\alpha\beta}$, after the nonlinear matrix operation process, become systematic error in $E_\alpha^{(1)}$. Furthermore, even if $E_\alpha^{(1)}$ is free of errors, other nonlinear matrix operations in finding the weight through Eq. \ref{eqn:W} will render other statistical errors into systematic errors. These systematic errors then lead to $\epsilon\neq 0$, which can be significant at large scales due to smaller modes and larger statistical fluctuations. This behavior (chance correlation) has been observed in CMB foreground removal with the internal linear combination (ILC) method (e.g. \citet{2008PhRvD..78b3003S,2008A&A...487..775V}). ILC is similar to our weighting scheme. It is also a linear estimator, with Eq. \ref{eqn:wq} replaced by the condition of unbiased CMB amplitude and Eq. \ref{eqn:ws} replaced by minimizing the variance of weighted maps. Given the resemblance in the math structure, we suspect that the $\epsilon<0$ at large scale found here has similar origins as that in ILC.   Since it is not the dominant source of systematic errors, we postpone its further investigation. On the other hand, $\epsilon\neq 0$ at $\ell\ga 10^3$ is expected from Fig. \ref{fig:eme}, since the scaling relation is no longer as accurate as the case of $\ell<10^3$. $\epsilon$ increases quickly with increasing $\ell$ and reaches $0.02$ at $\ell=3000$. This will be the dominant systematic error in the $\kappa$ reconstruction.

\begin{figure}
    \includegraphics[width=0.45\textwidth]{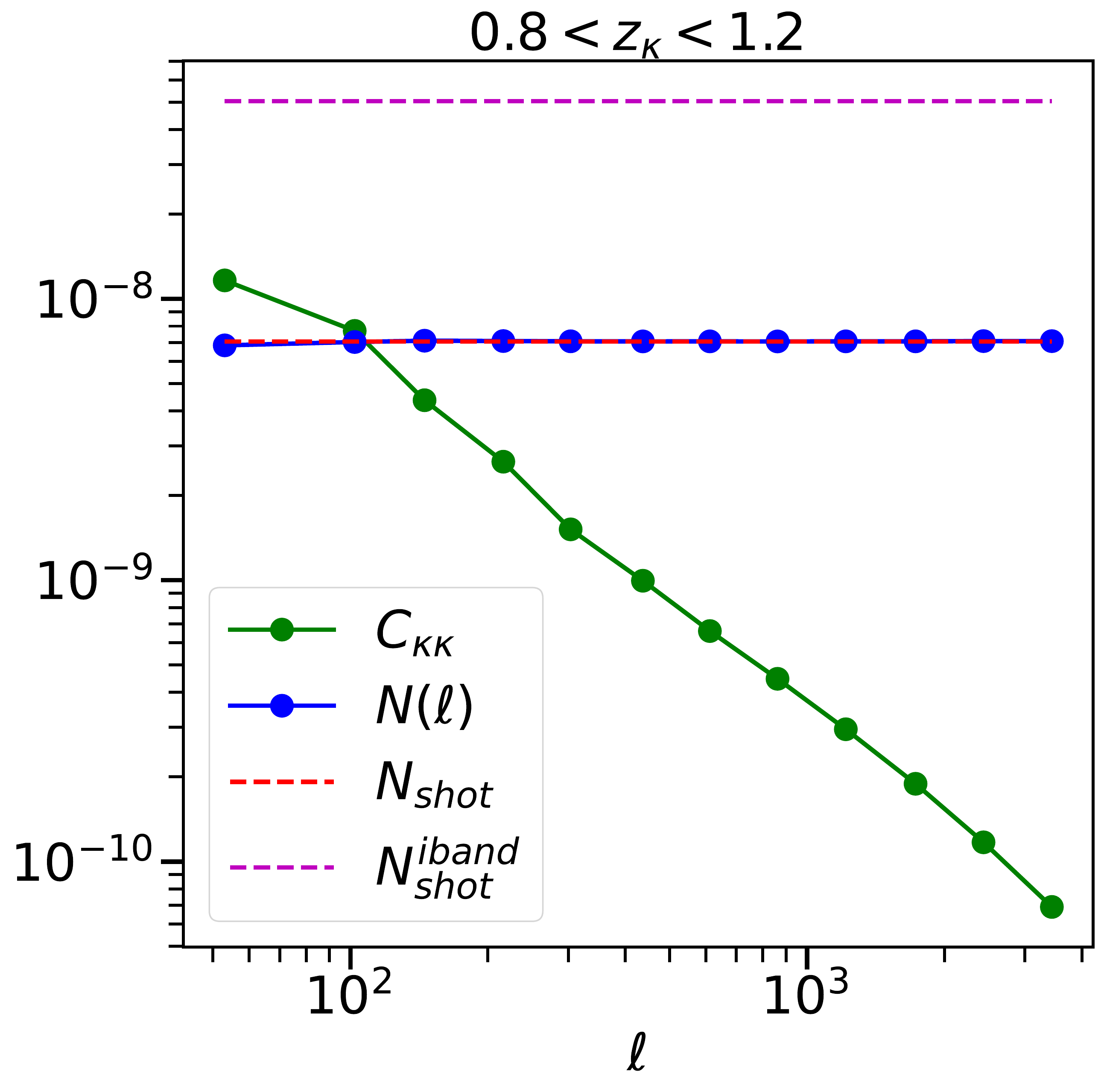}
	\caption{The reconstruction performance in the lensing auto power spectrum. The reconstructed map is signal dominated at $\ell<100$. The noise is dominated by the weighted shot noise ($N_{\rm shot}$). The weighting over multiple photometry bands heavily reduces the shot noise, comparing the case of using a single photometry band (e.g. $i$-band and $N^{i{\rm band}}_{\rm shot}$).
	\label{fig:Ckk-Nl}}
\end{figure}

\subsection{The auto-correlation error budget}

The map quality can be quantified by contaminations in the auto power spectrum,
\begin{equation}
    C_{\hat{\kappa}\hat{\kappa}}(\ell)=C_{\kappa\kappa}(\ell)+N(\ell)\ .
\end{equation}
The noise power spectrum $N(\ell)$ includes the auto power spectra of shot noise, $\sum_\alpha W_\alpha \delta^S_{g,\alpha}$ and $\epsilon\delta_m$. It also includes the $\kappa$-$\epsilon\delta_m$ cross term. To calculate $N(\ell)$, we calculate $C_{\hat{\kappa}\hat{\kappa}}$ directly from the reconstructed maps and $C_{\kappa\kappa}$ from the maps of the true signal. We then subtract $C_{\kappa\kappa}$ from $C_{\hat{\kappa}\hat{\kappa}}$ to obtain $N(\ell)$.   The resulting $N(\ell)$ is shown in Fig. \ref{fig:Ckk-Nl}. We also compare it to the weighted shot noise $N_{\rm shot}$, which is calculated using Eq. \ref{eqn:ws} given the weights.
Fig. \ref{fig:Ckk-Nl} shows that the error budget is dominated by $N_{\rm shot}$, namely $N(\ell)\simeq N_{\rm shot}(\ell)$. The reconstructed map is signal dominated at $\ell<100$, and is noise dominated at $\ell>200$.
The map is not as clean as the shear maps or CMB lensing maps, which will reach the cosmic variance limit at $\ell\la 1000$ (e.g. \cite{CMBS4}). The comparison here assumes the same number of galaxies for cosmic magnification and cosmic shear. But in principle there are more galaxies to use for cosmic magnification than for cosmic shear, since galaxy detection is easier than galaxy shape measurement. Therefore there are still rooms for improving the cosmic magnification reconstruction, by including more  galaxies at faint end.

One valuable lesson is to use as many photometry bands as possible. The noise power spectrum $N(\ell)$ when using $ugrizY$ photometry bands is a factor of $7$ smaller that of using only the $i$ band (Fig. \ref{fig:Ckk-Nl}). The relative ranking of galaxy brightness varies in different photometry bands. It has two-fold impacts on improving the reconstruction. First, galaxies have different $q=2(\alpha-1)$ in different photometry bands.  This source of cosmic magnification information  significantly improves the reconstruction. Second, shot noise in flux bins of different photometry bands is different. So combining all the flux bins reduce  shot noise.

\begin{figure}
    \centering
    \includegraphics[width=0.45\textwidth]{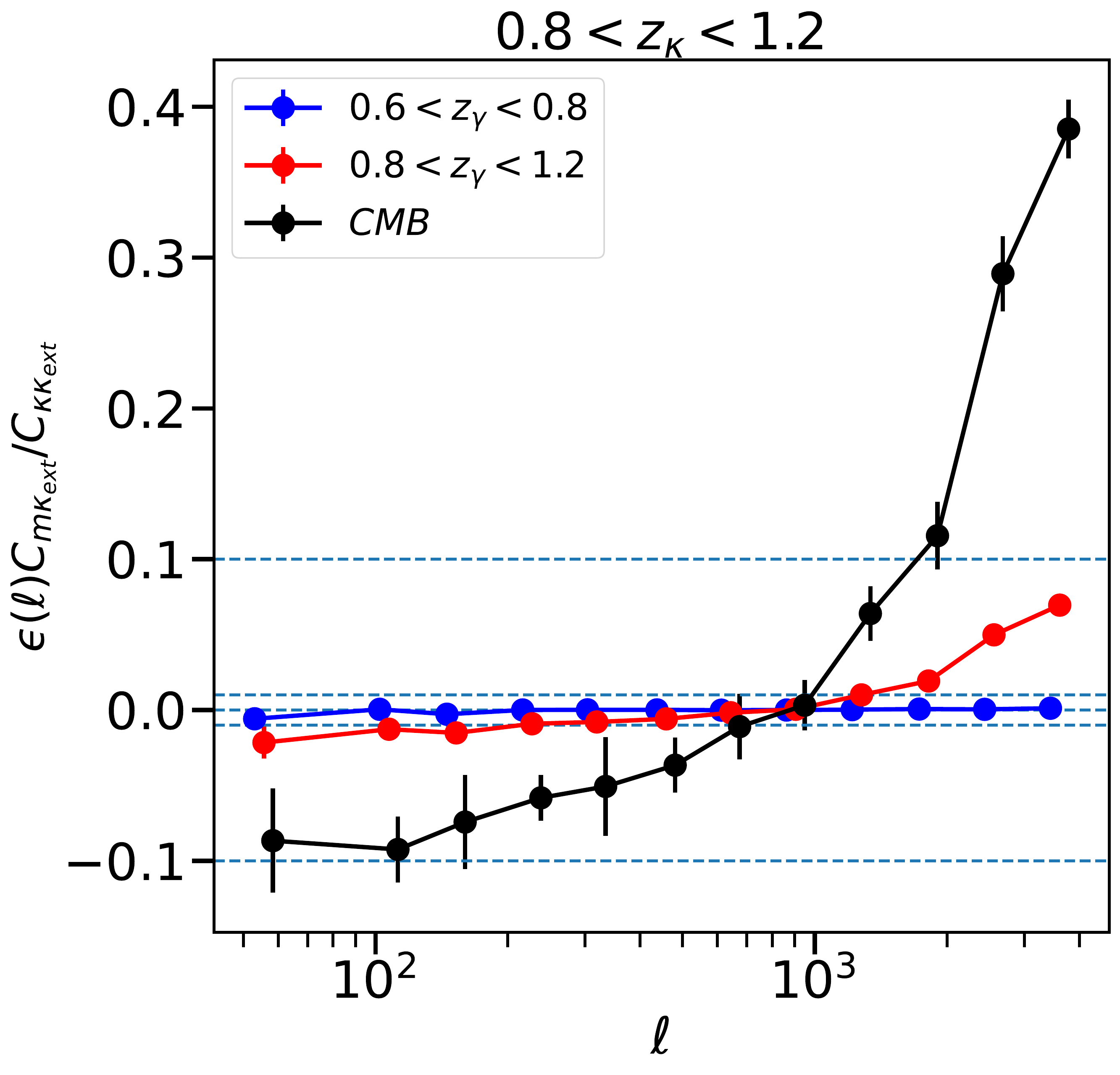}
    \caption{Systematic error in the $\hat{\kappa}$-external lensing cross-correlation measurements. We have investigated source redshifts $z_{\gamma}\in (0.2,2)$ and $z_{\rm CMB}$. The fractional error is negligible when $z_\gamma<z_\kappa$, so we only show the cases of $z_\gamma\in (0.6,0.8)$, $(0.8,1.2)$ and CMB lensing. The error increases with $z_\gamma$, but remains negligible at $\ell<10^3$. At $\ell\ga 2000$, the systematic error is significant, especially for cross-correlating with CMB lensing. }
    \label{fig:sys}
\end{figure}
\subsection{S/N estimation of cross-correlation with external fields}
Although the reconstructed $\kappa$ map will have large noise and the resulting auto power spectrum measurement S/N is hard to compete with cosmic shear or CMB lensing, the cross-correlation with external lensing fields is significantly cleaner. The reason is that  none of $\sum W_\alpha\delta_{g,\alpha}^S$ and shot noise biases the two-point statistics. Only the residual deterministic component of galaxy clustering induces bias in the cross-correlation measurement. The resulting two-point correlation function is
\begin{equation}
    C_{\hat{\kappa}\kappa_{\rm ext}}(\ell)=C_{\kappa\kappa_{\rm ext}}+\epsilon(\ell)C_{m\kappa_{\rm ext}}\ .
\end{equation}
Here $\kappa_{\rm ext}$ stands for the convergence field from external data such as cosmic shear and CMB lensing. This type of cross-correlation has been proposed and measured in HSC by \cite{2021PhRvD.103l3504L}.

 Since the matter distribution is only correlated with weak lensing at higher source redshift, $\epsilon\neq 0$  causes non-negligible contamination to $C_{\hat{\kappa}\kappa_{\rm ext}}$ only when the source redshifts of external lensing fields are higher than that of cosmic magnification, and only when $\ell\ga 2000$. Fig. \ref{fig:sys} shows the resulting errors in cross-correlation with cosmic shear and CMB lensing. For cosmic magnification reconstructed at $z_\kappa\sim 1$, its cross-correlation with cosmic shear at $z_\gamma<z_\kappa$ has negligible systematic error. Due to the redshift range and the extra broadening caused by photo-z errors, the systematic error becomes significant at $\ell>2000$ when $z_{\kappa,\gamma}\in (0.8,1.2)$. The error increases with increasing $z_\gamma$. So the situation is most severe for CMB lensing.

\begin{figure}
    \centering
    \includegraphics[width=0.45\textwidth]{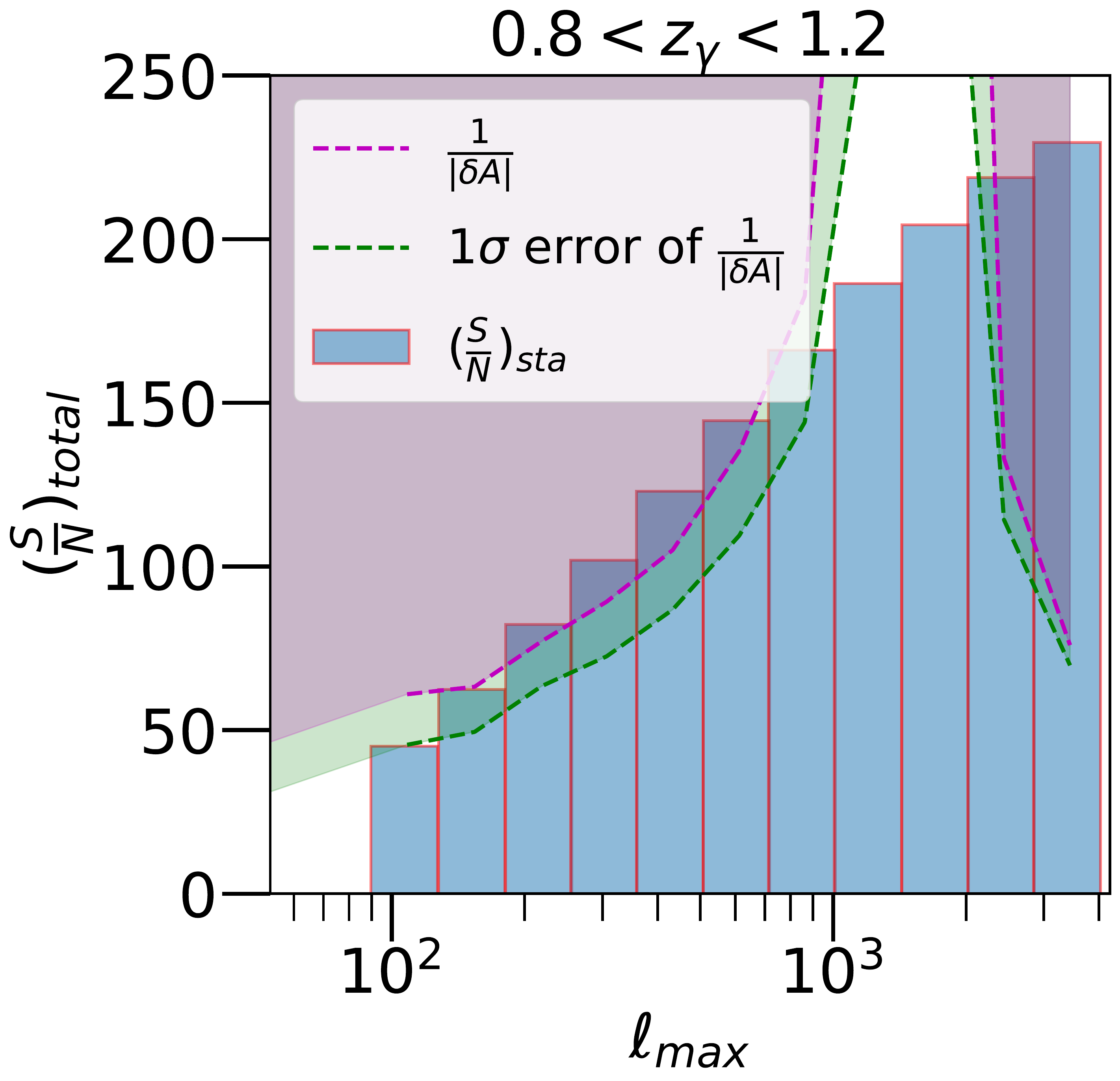}
	\caption{The S/N expected for measurements of cross-correlating the reconstructed convergence map with external shear at $0.8<z_\gamma<1.2$, for a LSST-like survey.  The histogram shows the S/N if only statistical errors are considered. Systematic errors are subdominant at $\ell<10^3$. \label{fig:SNR}}
\end{figure}

 The S/N of cross-correlation measurement, if only statistical error is considered,  can be estimated by
 \begin{equation}
 \label{eqn:GSN}
     \left(\frac{S}{N}\right)^2_{{\rm sta},\ell}=\frac{2\ell\Delta \ell f_{\rm sky}}{1+C_{\hat{\kappa}\hat{\kappa}}(C_{\kappa_{\rm ext}\kappa_{\rm ext}}+N_{\rm ext})/C^2_{\hat{\kappa}\kappa_{\rm ext}}}\ .
 \end{equation}
Here $\Delta \ell$ is the multipole $\ell$ bin size and $f_{\rm sky}$ is the sky coverage. We adopt $f_{\rm sky}=48\%$ \citep{LSST}. $N_{\rm ext}$ is the noise power spectrum of the external lensing field. This estimation does not consider super sample covariance \citep{2013PhRvD..87l3504T} or non-Gaussian covariance \citep{2009MNRAS.395.2065T}, which will decrease the S/N. Since the decrement is moderate (e.g. \citet{2023arXiv230404489Y}), we will adopt this simplification of Eq. \ref{eqn:GSN} to demonstrate the potential of cosmic magnification reconstruction. The estimated S/N is shown in Fig. \ref{fig:SNR}.

However, the above raw estimation of S/N is not realistic, due to the systematic error, which puts an upper limit on the realistic S/N. Since it varies with scale, we quantify its impact on the overall amplitude of the measured cross power spectrum. Namely we try to fit the data $C_{\hat{\kappa}\kappa_{\rm ext}}$ with $AC_{\kappa \kappa_{\rm ext}}$. $A=1$ means vanishing weighted systematic error. The bias in $A$ is then
\begin{equation}
\label{eqn:deltaA}
    \delta A=\frac{\sum_{\ell} (\epsilon C_{m\kappa}/C_{\kappa \kappa_{\rm ext}}) (S/N)^2_{{\rm sta},\ell}}{\sum_{\ell} (S/N)^2_{{\rm sta},\ell}}\ .
\end{equation}
Therefore a realistic estimation of S/N is
\begin{equation}
    \frac{S}{N}={\rm min}\left(\left(\frac{S}{N}\right)_{{\rm sta}}\ ,\ \frac{1}{|\delta A|} \right)\ .
\end{equation}
Fig. \ref{fig:SNR} shows the S/N expected by cross-correlating with cosmic shear at $0.8<z_\gamma<1.2$, for a LSST-like survey. Cross-correlation measurement with  $S/N\sim 200$  at $\ell<2000$ can be achieved. Fig. \ref{fig:sys-sta} shows  the statistical and systematic errors for multipole $z_\gamma$ bins.

Systematic errors, in particular at $\ell>3000$, can be further mitigated. Fig. \ref{fig:sys-sta} shows more detailed comparison between the statistical and systematic errors, for multipole $z_\gamma$ bins. It shows that two behave differently in the $\ell$ space. For example, the raw measurement of $C_{\hat{\kappa}\kappa_{\rm ext}}$ can be fitted by $C_{\kappa\kappa_{\rm ext}}+\epsilon C_{m\kappa}$ of multiple $z_\gamma$ bins. Since $\epsilon$ does not vary with $z_\gamma$, we can break the degeneracy between $C_{\kappa\kappa_{\rm ext}}$ and $\epsilon C_{m\kappa}$ with multiple $z_\gamma$ (cosmic shear redshift ranges). We will leave such investigations until further study.

\section{Conclusion}
\label{sec:conclusion}
We propose a minimum variance linear estimator to reconstruct the weak lensing convergence map, by weighing overdensity maps of galaxies in different magnitude bins and different photometry bands. This estimator efficiently suppresses the major contamination of galaxy intrinsic clustering. In particular, it is able to suppress the deterministic part of the galaxy intrinsic clustering by a factor of $10^2$ (Fig. \ref{fig:epsilon}). Therefore the reconstructed map is nearly free of galaxy clustering contamination in its cross correlation with cosmic shear at $\ell\la 2000$. Based on the CosmoDC2 galaxy mock and the CosmicGrowth simulation, we validate this method. We also forecast that a LSST-like survey can achieve $S/N\ga 200$ through cross-correlations between the reconstructed convergence map and cosmic shear, with systematic errors under control.

There are important issues for further investigation. (1) One is to further suppress the residual galaxy clustering in the convergence-shear cross-correlation measurement.  Such systematic error becomes dominant at $\ell\ga 10^3$ and when the shear source redshift $z_\gamma$ is higher than the convergence source redshift $z_\kappa$ (Fig. \ref{fig:sys-sta}). We believe that it can be further reduced, since we know that it is proportional to the matter-cosmic shear cross-correlation $C_{m\kappa_{\rm ext}}$. The only unknown  parameter is its amplitude $\epsilon$. $\epsilon$ varies with $\ell$ (Fig. \ref{fig:epsilon}), but not $z_\gamma$. Therefore by joint analysis of cross-correlations of multiple $z_\gamma$, we can
simultaneously constrain cosmological parameters and $\epsilon(\ell)$. (2) Although the reconstructed lensing convergence map can deliver high S/N cross-correlation measurement with cosmic shear and CMB lensing, auto power spectrum measurement in the current investigation is not competitive with respect to cosmic shear and CMB lensing (Fig. \ref{fig:Ckk-Nl}). However, there are still possibilities to improve. The lesson learn from the current investigation is to weigh over as many galaxy overdensity maps as possible. In this work we use $26$ maps defined by magnitude cut in the $ugrizY$ photometry bands. \citet{2023arXiv230411540Z} found that galaxy sub-samples defined in the color-space are also useful in understanding and controlling galaxy bias (clustering). Since gravitational lensing does not change color, the cosmic magnification (magnification bias) in the color bins has $q=-2$ when we ignore any other selection effects.  This means that including galaxy sub-samples in the color space and the associated overdensity map, the lensing reconstruction can be further improved, in both S/N and systematic errors.

\section*{Acknowledgement}
This work is supported by National Science Foundation of China (11621303, 12273020, 11890690), the National Key R\&D Program of China (2020YFC2201602), the
China Manned Space Project (\#CMS-CSST-2021-A02 \& CMS-CSST-2021-A03), and the Fundamental Research Funds for the Central Universities.  YY thanks the sponsorship from Yangyang Development Fund.

\section*{Data Availability}
The data underlying this article will be shared on reasonable request to the corresponding author.


\bibliographystyle{mnras}
\bibliography{mybib}

\appendix
\section{The mocks for the tests}
\label{sec:mocks}
\begin{figure}
\centering
    \includegraphics[width=0.4\textwidth]{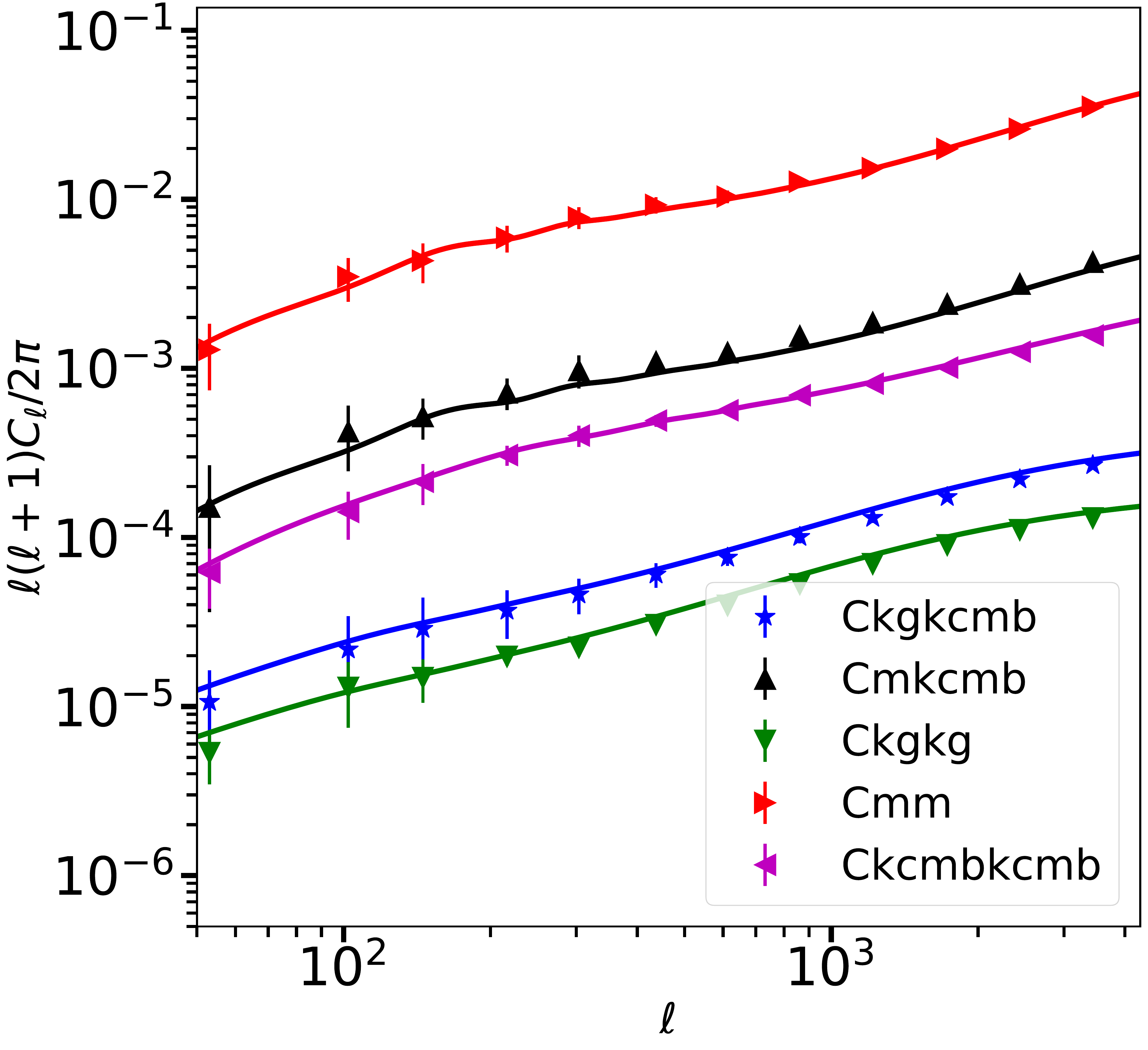}
    \caption{The angular power spectra of the simulated galaxy maps, matter maps and lensing maps. All points and errorbars are from the average of 20 mocks. These simulated power spectra agree with the theoretical prediction of CAMB (solid curves). \label{fig:LIM}}
\end{figure}
We use CosmoDC2 to test our method. CosmoDC2 is the galaxy mock generated for LSST \citep{CosmoDC2}. It provides a lightcone of $440\ \mathrm{deg}^2$, with $\sim 2.26$ billion galaxies to $z=3$. The magnitudes in $ugrizY$ photometry bands  are also provided. Such  multi-band photometry information is crucial for our method.  CosmoDC2 also provides the host halo information of these galaxies. A problem is that, CosmoDC2 itself does not provide the associated 3D matter density field, which is based on the Outer Rim simulation \citep{OuterRim}. Since we need the underlying density field for our analysis, we combine CosmoDC2 with our CosmicGrowth simulations \citep{Jing19}. The CosmicGrowth simulation we use has a flat $\Lambda$CDM cosmology with $\Omega_m=0.268$, $\Omega_b=0.0445$, $\sigma_8=0.83$, $h=0.71$ \& $n_s=0.968$. The simulation has box size of $L_{\rm box}=1200 h^{-1}\ \mathrm{Mpc}$, and particle number of $3072^3$.   We use the galaxy-host halo information of CosmoDC2 to assign galaxies to the CosmicGrowth simulation halos.  The bottom line is to  have the same amount of galaxies about the same magnitude in both simulations. Since CosmoDC2 only contains halos hosting galaxies, we compare the CosmoDC2 halo mass function with the halo mass function  predicted by the CosmoDC2 cosmology to figure out the fraction of halos of given mass hosting CosmoDC2 galaxies. Since the cosmological parameters in the two simulations are not identical, the two have different halo mass functions. We first do an abundance matching between the halo mass function predicted by the CosmoDC2 cosmology and that by the CosmicGrowth cosmology.  This tells us the mass of halos in CosmicGrowth corresonding to that in CosmoDC2. Then we can match each halo in CosmoDC2 with the corresponding halo (or one of the corresponding halos) in CosmicGrowth, we transport all galaxies in the CosmoDC2 halo to the CosmicGrowth halo, including the distinction of central and satellite galaxies and their relative positions. We caution that this procedure of transporting CosmoDC2 galaxies to other simulations is only approximate, since properties other than the halo mass (e.g. halo  ellipticity, environment and history) also affect galaxy formation. In the future we should use more realistic galaxy mocks to test our method.

We generate 20 light-cones of 3D galaxy positions and photometry information to $z=2.48$, each of size $8.2^{\circ}\times 8.2^{\circ}$.  We then add random redshift scatter of r.m.s. $\sigma_z=0.05(1+z)$ to mock galaxies.  Magnitude cuts are applied to define galaxy sub-samples for the galaxies with $0.8<z^P<1.2$, as shown in Table \ref{tab:band}.  These galaxy sub-samples are used to generate maps of (unlensed) galaxy overdensity on $512^2$ uniform grids. We also generate corresponding light cones of matter distribution and weak lensing map at various source redshifts (including $z_s=1100$ for CMB lensing).   We use Eq. \ref{eqn:deltaL} to combine the galaxy ovendensity maps and the lensing convergence maps into the lensed galaxy overdensity maps, with $q$ measured from the luminosity function of mock galaxies.
\begin{figure}
\centering
   \includegraphics[width=0.4\textwidth]{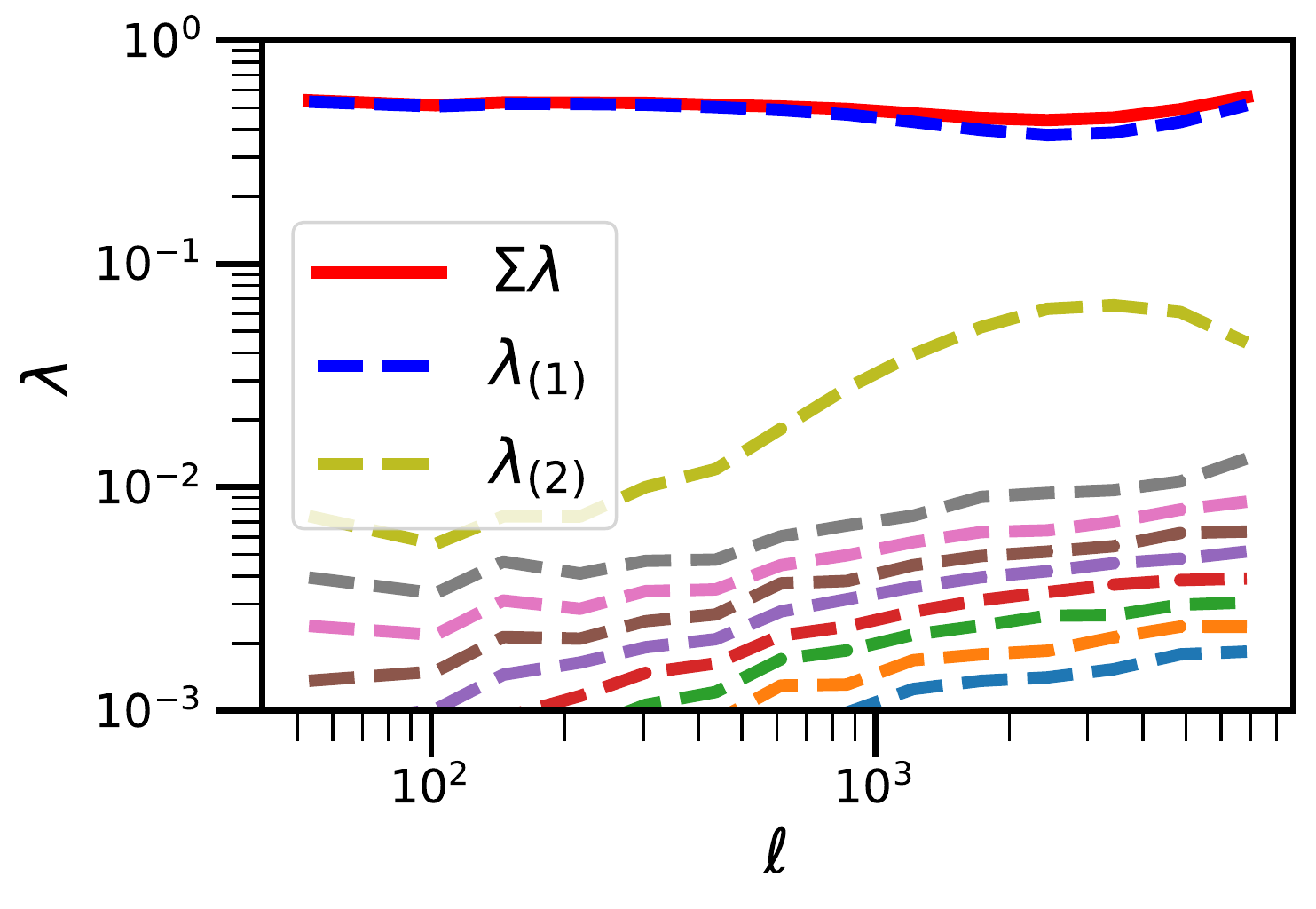}
    \caption{The eigenvalue of the matrix $C_{\alpha\beta}$ of galaxies at $0.8<z_\kappa<1.2$. The dashed lines show different eigenvalues and the solid line is the sum of eigenvalues. The first eigenmode overwhelms the others. The second eigenmode is statistically significant, while the others are insignificant comparing to shot noise fluctuations. Our analysis only uses the first (dominant/principal) eigenmode, so it is robust to potential numerical errors in the matrix operation.  \label{fig:EV}}
\end{figure}

We check the validity of the generated maps by measuring the corresponding power spectra.  Fig. \ref{fig:LIM} shows the measurements of $C_{\kappa\kappa}$ for the source redshift $0.8<z^P<1.2$. It agrees with the theoretical prediction via CAMB \citep{Lewis1999EfficientCO}. It also shows $C_{mm}$. Besides the correct auto-correlation statistics, the maps also have correct cross-correlations, such as the matter(galaxy)-CMB lensing cross-correlation $C_{m\kappa_{\rm CMB}}$ and galaxy lensing-CMB lensing cross-correlation $C_{\kappa_g\kappa_{\rm CMB}}$.

We then measure the galaxy correlation matrix $C_{\alpha\beta}$ ($\alpha,\beta=1,\cdots,N$). Here $N=26$ is for our fiducial galaxy sample. Notice that $C_{\alpha\beta}$ also include contributions from cosmic magnification.  The eigenvalues, as a function of $\ell$, are shown in Fig. \ref{fig:EV}. The matrix is dominated by the largest eigenmode, while the second largest eigenmode is also statistically significant. These behaviors are consistent with previous findings \citep{Bonoli09,Hamaus10}.

We test the proposed proportionality relation between the largest eigenvector $E^{(1)}_\alpha$ and the galaxy deterministic bias $b_{D,\alpha}$ (Eq. \ref{eqn:be}). As a reminder, $\alpha$ denotes the $\alpha$-th galaxy sub-sample.
Fig. \ref{fig:eme} shows that this relation holds accurately at $\ell<1000$ and reasonably well at $1000<\ell\la 3000$. This is consistent with our finding in 3D galaxy clustering \citep{2023arXiv230411540Z}. Results present here show that the proportionality relation is robust against mixing of different $k$ mode due to the projection effect in angular clustering. Therefore we are able to use Eq. \ref{eqn:be} to construct the optimal weight of reconstructing the lensing convergence maps. Notice that for the estimator, the proportionality coefficient does not matter.

\begin{figure}
    \centering
   \includegraphics[width=0.4\textwidth]{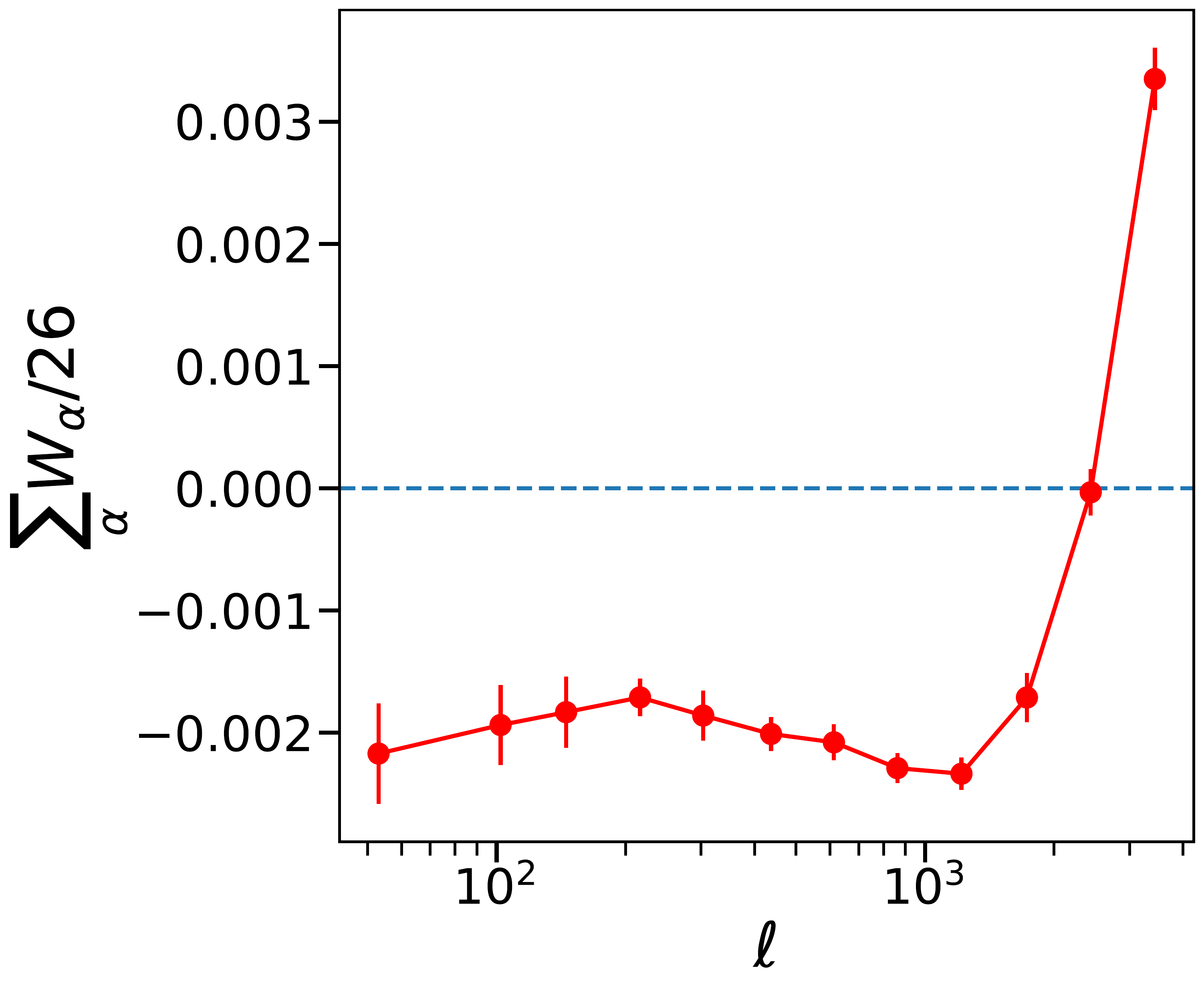}
    \caption{The mean $W$, defined by $\sum_{\alpha} W_{\alpha}/\sum_\alpha$ ($\alpha=1,2\cdots 26$). We have avearged over the 20 maps. It is a factor of $10$ smaller than the typical value of $|W_\alpha|\sim 0.1$. This results shows that the condition $\sum_\alpha W_\alpha=0$ required by \citet{2021RAA....21..247H} is approximately satisfied. }
    \label{fig:sumwi}
\end{figure}

Fig. \ref{fig:sumwi} shows an interesting result that $\langle W\rangle\equiv \sum_\alpha W\alpha/\sum_\alpha\sim 0$. More exactly, $\langle W\rangle$ is a factor of $\sim 10$ smaller than typical $W_\alpha$ (Fig. \ref{fig:weight}). This means that in reality the condition $\sum_\alpha W_\alpha=0$ adopted by the previous proposal \citep{2021RAA....21..247H} is reasonably satisfied. For this reason, we may adopt the condition $\sum_\alpha W_\alpha=0$ as a first attempt to analyze real data.\footnote{Jian Qin, et al, in preparation.} This version of lensing convergence reconstruction, although sub-optimal in suppressing $\epsilon$, is applicable to galaxy catalogs without accurate galaxy clustering measurement.


\section{Statistical and systematic errors}
\label{sec:errors}

\subsection{The reconstructed maps}
\label{subsec:maps}
\begin{figure}
    \centering
    \includegraphics[width=0.45\textwidth]{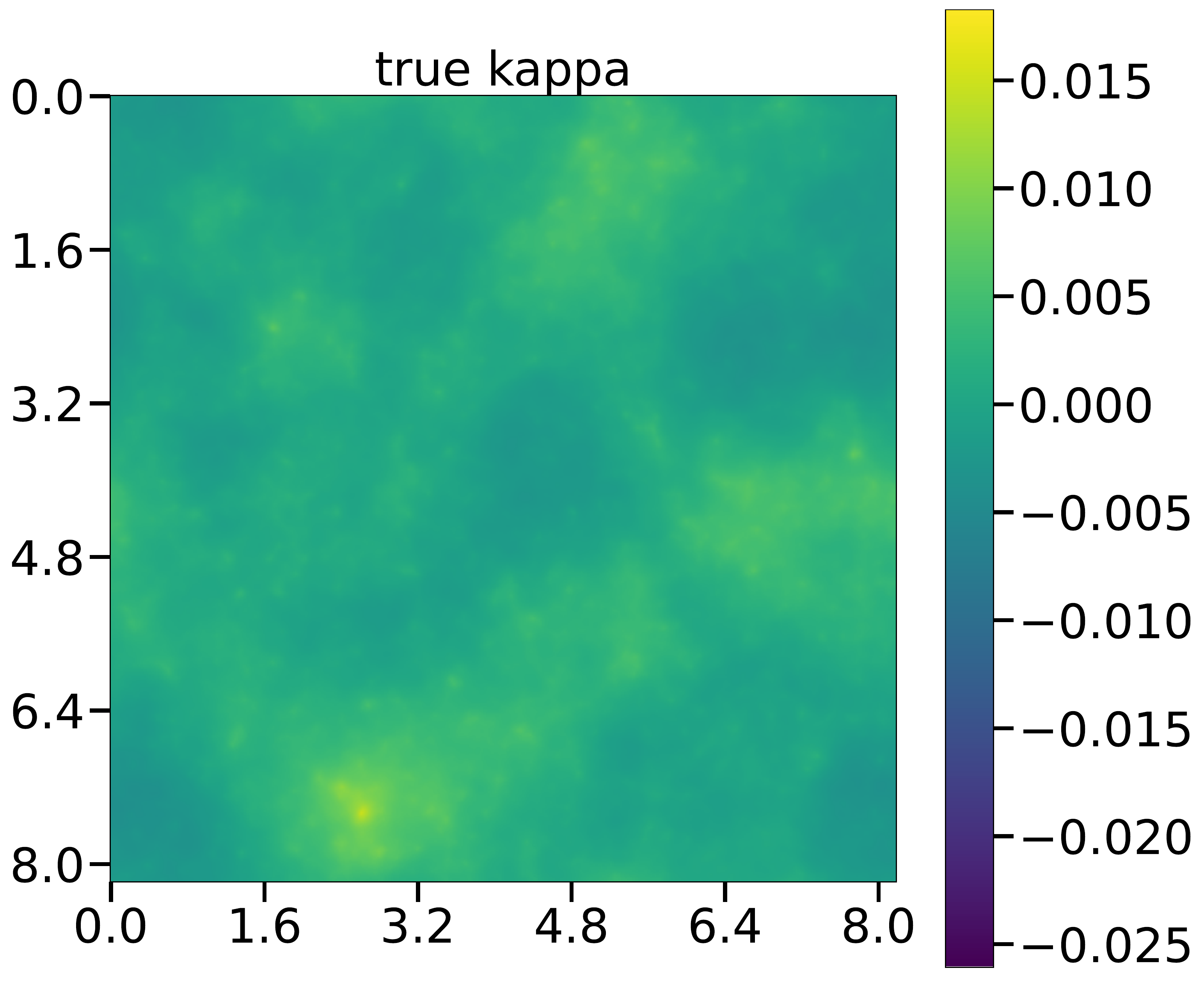}
    \includegraphics[width=0.45\textwidth]{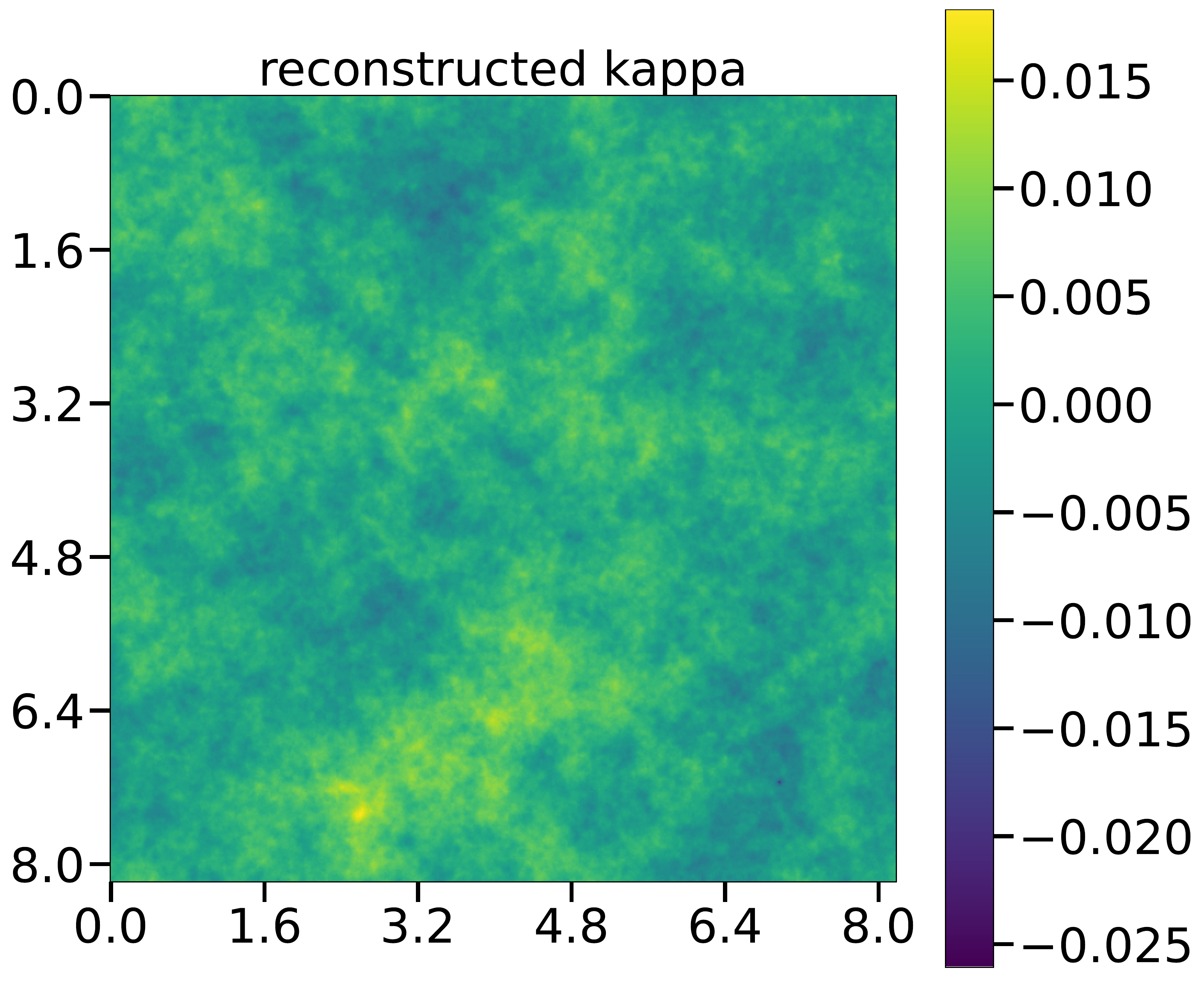}
	\caption{The true $\kappa$ map (top panel) and the reconstructed $\hat{\kappa}$ map (bottom panel), for one of the simulation realizations. Both maps are Wiener filtered by $W_{\rm WF}(\ell)=C_{\kappa\kappa}/(C_{\kappa\kappa}+N)$. The reconstructed map is noisy and shot noise in galaxy number count overwhelms at sub-degree scales. Nevertheless, the two maps show similarity at degree scales.  \label{fig:kappafield}}
\end{figure}

Fig. \ref{fig:kappafield} shows one of the reconstructed $\kappa$ maps.  The reconstruction is noisy, due to shot noise overwhelming at most scales. For this reason, we only present the Wiener filtered maps with $W_{\rm WF}=C_{\kappa\kappa}(\ell)/(C_{\kappa\kappa}(\ell)+N(\ell))$, in which both the lensing power spectrum $C_{\kappa\kappa}$ and the noise (mostly shot noise) power spectrum $N(\ell)$ are defined in the main text.  The process of Wiener filter reveals similarity of $\hat{\kappa}$ and $\kappa$ at degree scale and above. This is consistent with Fig. \ref{fig:Ckk-Nl}.

\subsection{The raw SNR}
\begin{figure}
    \includegraphics[width=0.24\textwidth]{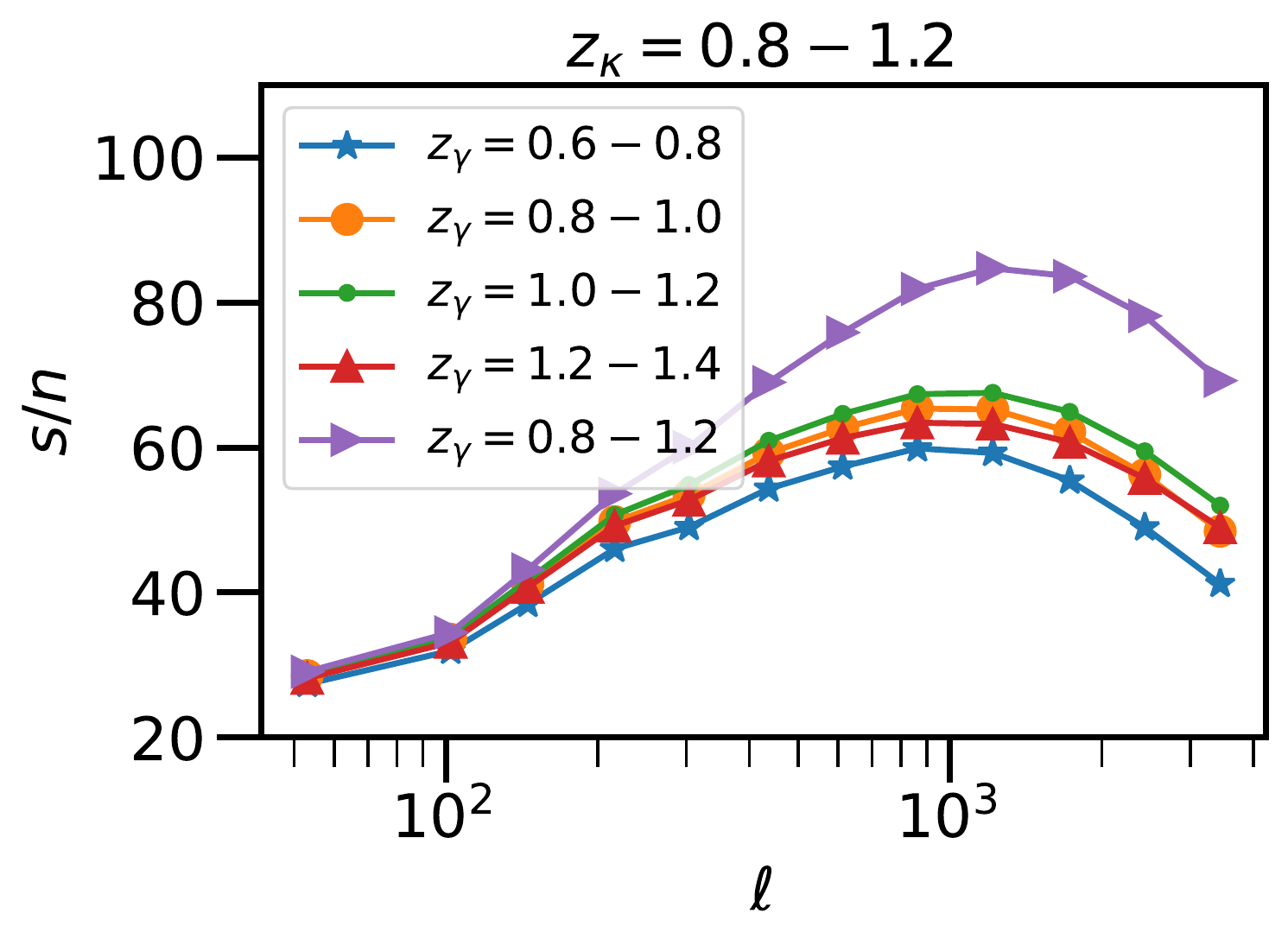}
    \includegraphics[width=0.24\textwidth]{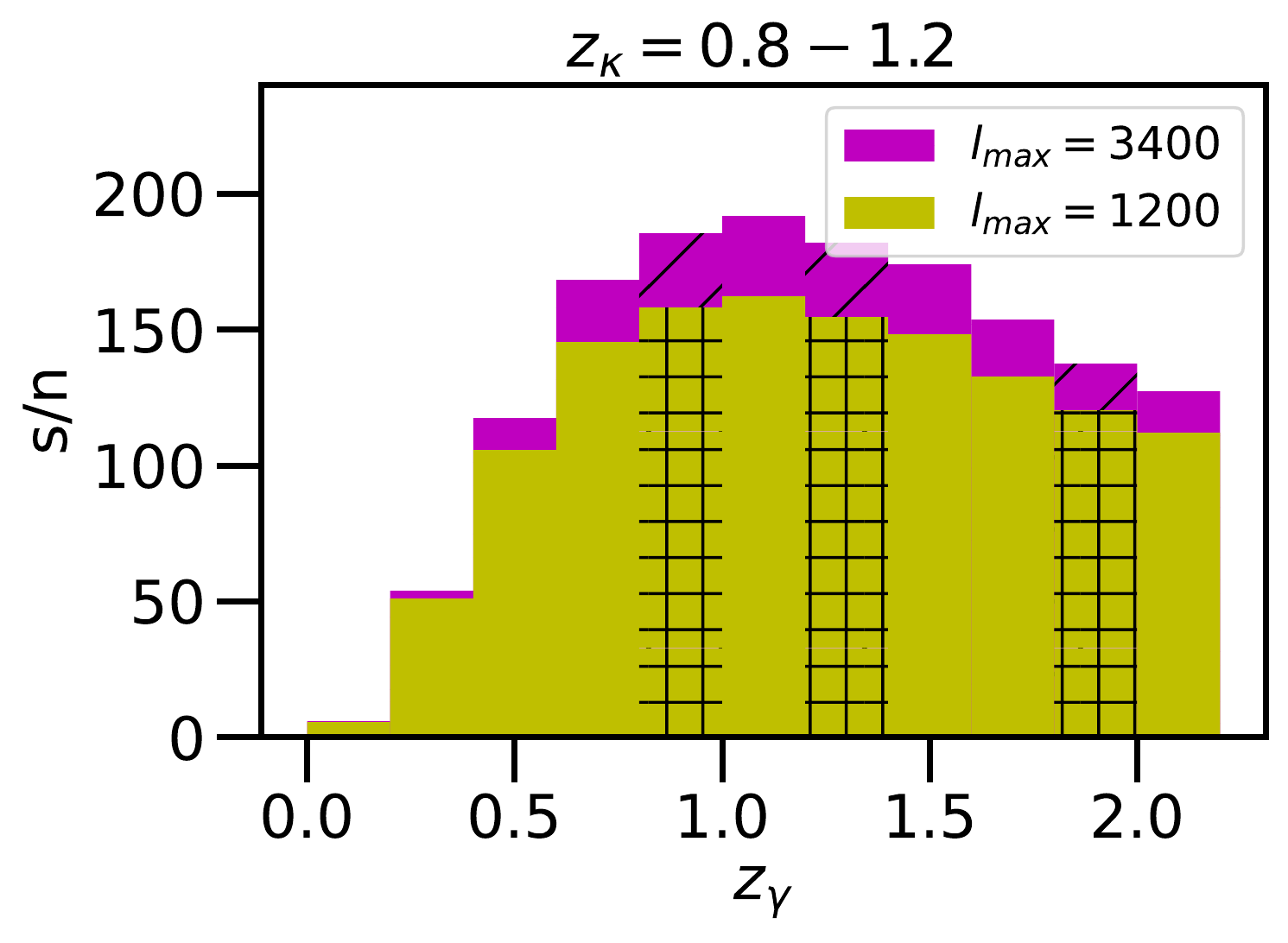}
	\caption{The raw SNR of cross-correlation between the reconstructed lensing maps and cosmic shear. Left panels  show the SNR at each $\ell$ bin, but for cross-correlating with cosmic shear at different redshifts $z_\gamma$ as labelled. Right panels show the total SNR up to $\ell=1200$ (3400), as a function of $z_\gamma$ of $\Delta z_\gamma=0.2$ redshift width. The total SNR can reach above $10^2$ for multiple $z_\gamma$.  \label{fig:SNR2}}
\end{figure}

The raw SNR, namely the signal-to-noise ratio only considering statistical errors, is shown in Fig. \ref{fig:SNR2}. The bottom left panel shows the SNR of each $\ell$ bin, for fixed convergence map constructed using galaxies at $0.8<z_\kappa<1.2$. The cross correlation is done with cosmic shear at various source redshifts $z_\gamma$. The peak contribution occurs at $\ell\sim 10^3$, where statistical errors from cosmic variance and shot noise roughly balance. The bottom right panel shows the total SNR up to $\ell_{\rm max}=1000$ or $3000$, for various $z_\gamma$. The maximum SNR is achieved when $z_\gamma\sim 1$, where the cross-correlation coefficient of the two lensing fields (cosmic convergence and shear) is the largest. For $z_\gamma\sim z_\kappa\sim 1$, the total SNR using all $\ell<10^3$ modes is about $200$.

\subsection{Systematic error versus statistical error}
The raw SNR is the optimistic case where systematic errors are ignored. As Fig. \ref{fig:epsilon} \& \ref{fig:sys} show, the systematic error due to imperfect correction of deterministic galaxy bias presents at all $\ell$. Its impact on the cross-correlation measurement is negligible when the cosmic shear redshift $z_\gamma<z_\kappa$.  The reason is that this contamination is induced by the residual deterministic galaxy bias and is therefore uncorrelated with cosmic shear at lower redshifts. But when $z_\gamma \ga z_\kappa$, this contamination will become more and more significant (Fig. \ref{fig:sys}).

\begin{figure}
    \includegraphics[width=0.24\textwidth]{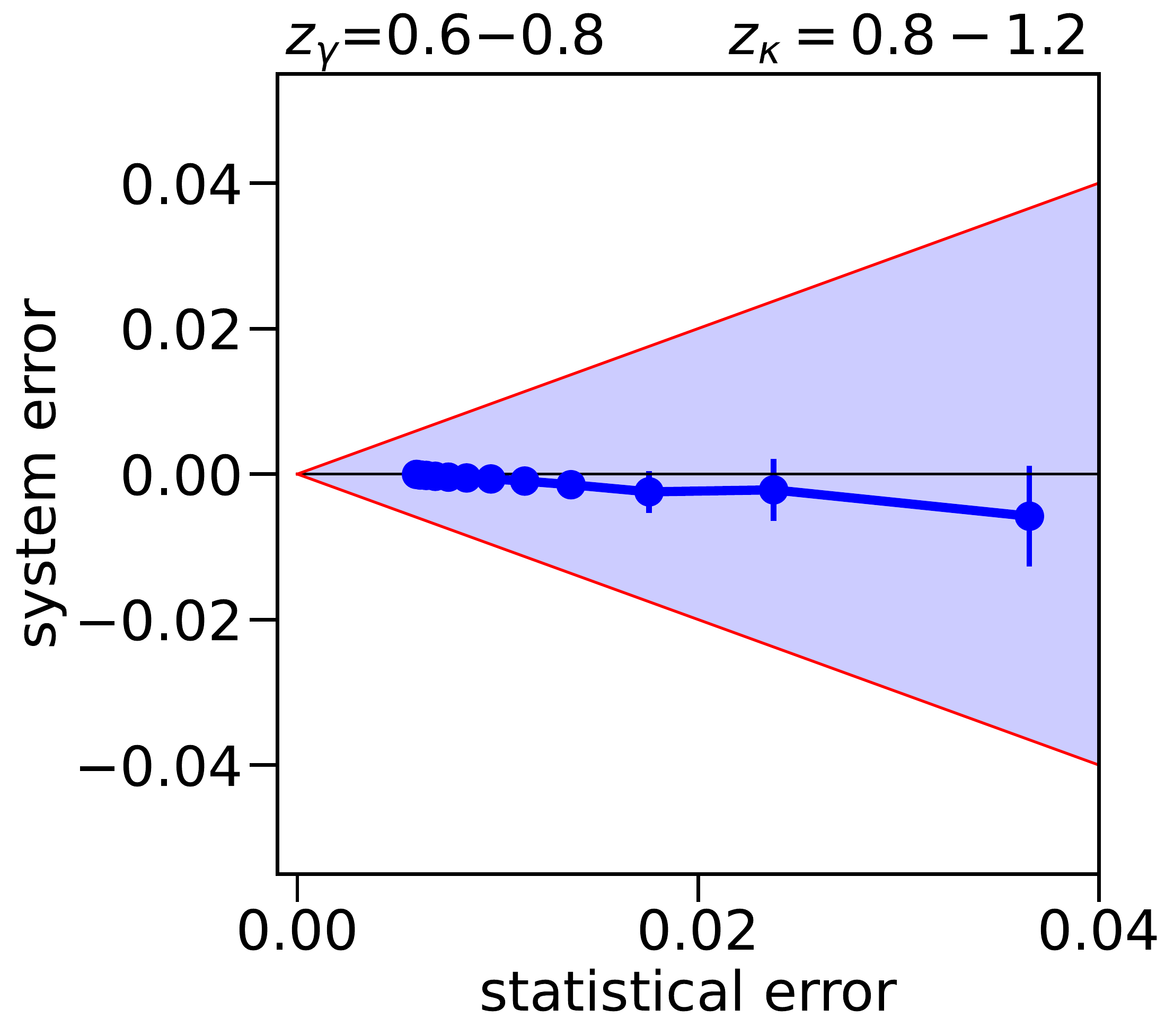}
    \includegraphics[width=0.24\textwidth]{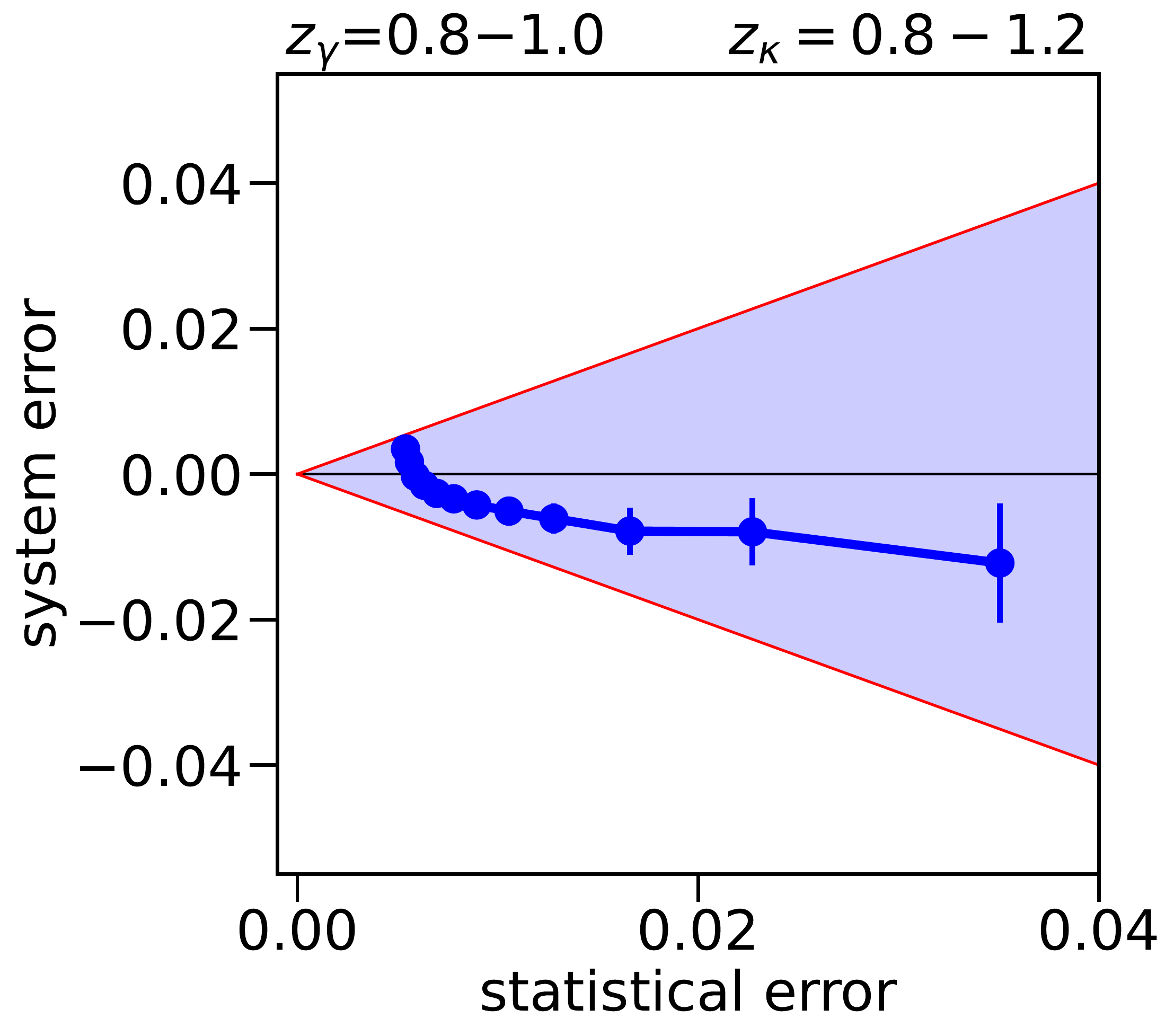}
    \includegraphics[width=0.24\textwidth]{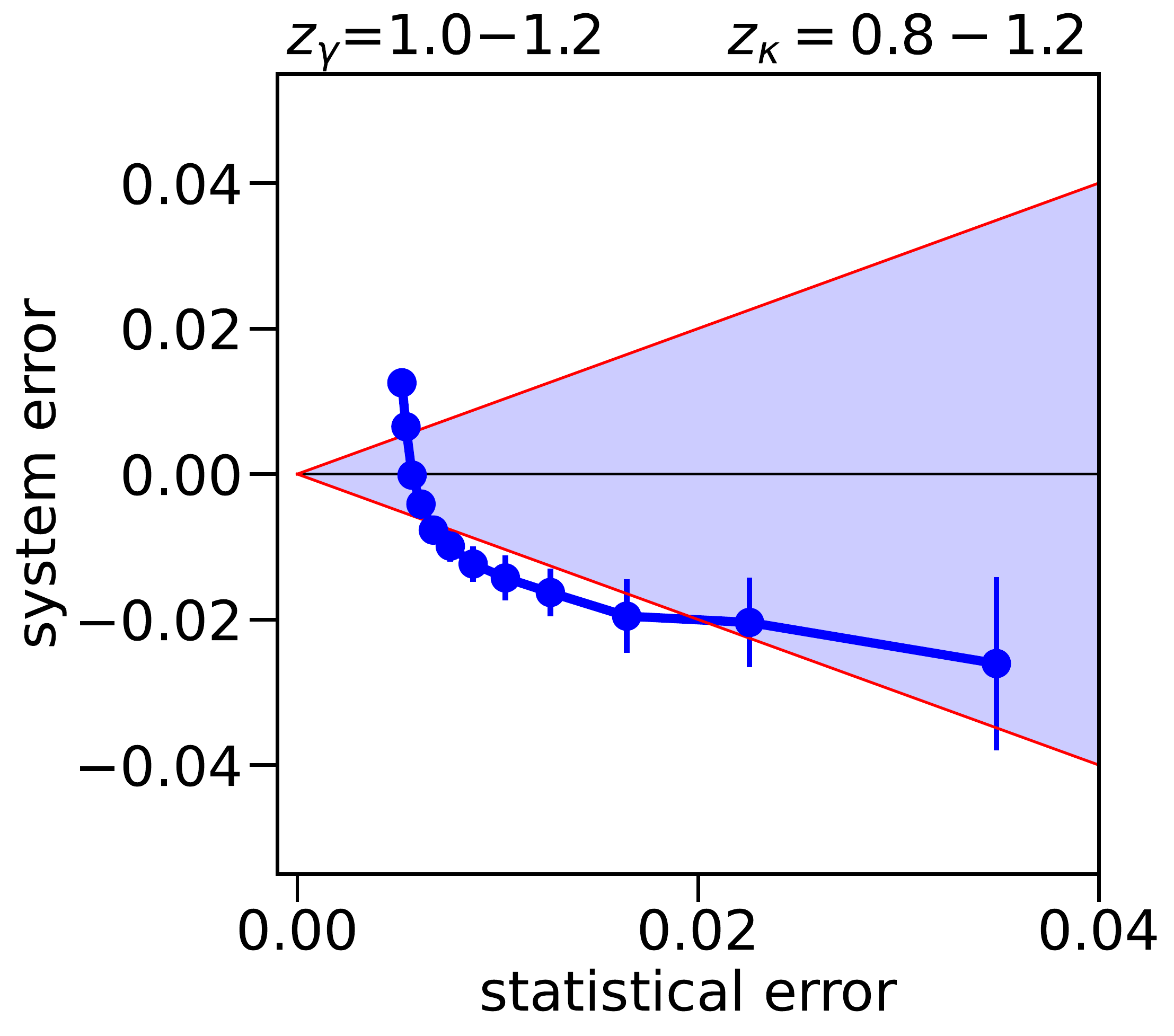}
    \includegraphics[width=0.24\textwidth]{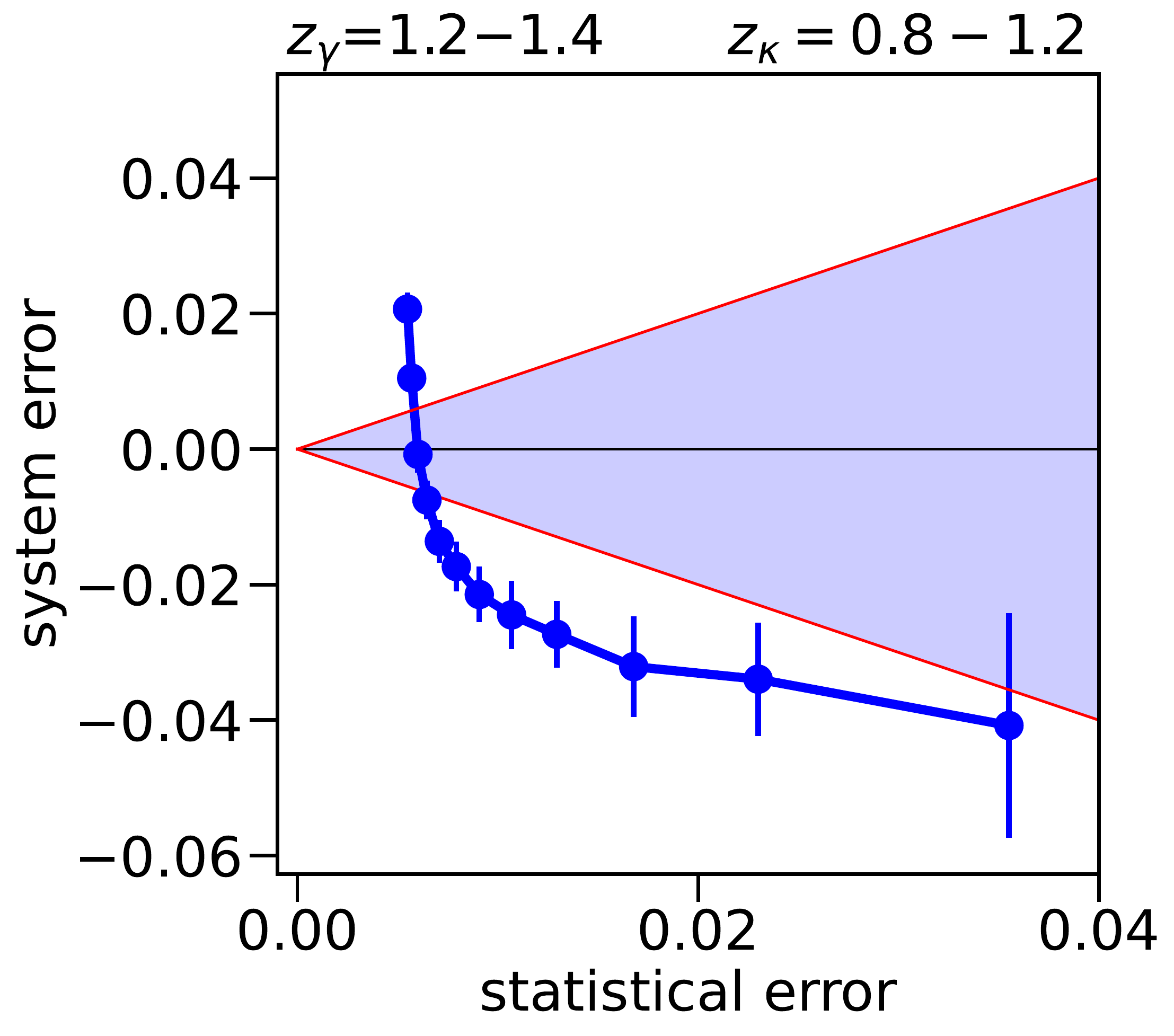}
	\caption{The systemic error versus statistical error plot. Each data points correspond to a $\ell_{\rm max}$, and a larger $\ell_{\rm max}$ in each panel corresponds to a smaller statistical error. The error bars are the r.m.s. of systematic errors, estimated using 20 realizations. Systematic error is subdominant to statistical error in the shaded region. Only when $z_\gamma>z_\kappa$ and only when $\ell\ga 10^3$, systematic errors may become statistically significant.  \label{fig:sys-sta}}
\end{figure}

Since both the amplitude and the sign of this systematic error varies with the angular scale, we adopt a commonly used weighting $(S/N)^2$ to weigh its overall impact at $\ell<\ell_{\rm max}$ and compress the overall systematic error into a single parameter $\delta A$ (Eq. \ref{eqn:deltaA}).  Fig. \ref{fig:sys-sta} shows the plot of the systematic error ($\delta A$) versus the statistical error ($1/(S/N)_{\rm stat}$). For brevity, we only show the result for the lensing convergence map reconstructed using galaxies in the photometric redshift range $0.8<z_\kappa<1.2$. The cosmic shear source redshifts are from $z_\gamma\in (0.6,0.8)$ to $z_\gamma\in (1.2,1.4)$. As expected, since the residual galaxy clustering only correlates with cosmic shear at higher redshifts, its contamination to the cross-correlation measurement with lower redshift (e.g. $z_\gamma\in (0.6,0.8)$) cosmic shear is negligible.\footnote{Notice that in such case the contamination is non-zero due to photo-z errors. } For this reason, we do not show the results at even lower cosmic shear source redshift. When $z_{\gamma}\in (1.0,1.2)$, the systematic error begins to dominate over the statistical error when the raw S/N reaches $\sim 200$ and prohibits more accurate measurement. The situation becomes worse for $z_\gamma\in (1.2,1.4)$ and even higher redshifts.

Mitigating these significant systematic errors will be a remaining major issue. As we have discussed in \S \ref{sec:conclusion}, there are promising approaches to correct such errors efficiently, either with a joint analysis of multipole $z_\gamma$ cross-correlation measurements or with $\kappa$ reconstruction of galaxy sub-samples in color space. These will be major tasks of future investigation.

\end{document}